\documentclass[paper]{JHEP3}

\usepackage{cite}
\usepackage{epsfig}
\usepackage{amsfonts}
\usepackage{amsmath}
\usepackage{amssymb}

\def \beq {\begin{equation}}
\def \eeq {\end{equation}}
\def \bea {\begin{eqnarray}}
\def \eea {\end{eqnarray}}

\def\Z#1{_{\lower2pt\hbox{$\scriptstyle#1$}}}
\def\X#1{_{\lower2pt\hbox{$\scriptscriptstyle#1$}}}

\title{Entropy Bound and Causality Violation in Higher Curvature Gravity}

\author{Ishwaree P. Neupane\\
Department of Physics and Astronomy, University of Canterbury\\
Private Bag 4800, Christchurch 8020, New Zealand
\\\email{E-mail:ishwaree.neupane@canterbury.ac.nz}}

\author{Naresh Dadhich\\
Inter-University Centre for Astronomy and Astrophysics, Pune 411
007, India \\\email{E-mail: nkd@iucaa.ernet.in}}

\abstract{In any quantum theory of gravity we do expect
corrections to Einstein gravity to occur. Yet, at fundamental
level, it is not apparent what the most relevant corrections are.
We argue that the generic curvature square corrections present in
lower dimensional actions of various compactified string theories
provide a natural passage between the classical and quantum realms
of gravity. The Gauss-Bonnet and $({\rm Riemann})^2$ gravities, in
particular, provide concrete examples in which inconsistency of a
theory, such as, a violation of microcausality, and a classical
limit on black hole entropy are correlated. In such theories the
ratio of the shear viscosity to the entropy density, $\eta/s$, can
be smaller than for a boundary conformal field theory with
Einstein gravity dual. This result is interesting from the
viewpoint that the nuclear matter or quark-gluon plasma produced
(such as at RHIC) under extreme densities and temperatures may
violate the conjectured KSS bound $\eta/s\ge 1/4\pi$, {\it albeit}
marginally so. }


\preprint{arXiv:0808.1919}

\begin{document}

\section{The Problem of Universality}

Black holes are thermal objects and hence they radiate as pure
black bodies with characteristic thermodynamic properties such as
temperature and entropy. In pure Einstein gravity, all black holes
do satisfy the famous Bekenstein-Hawking entropy
law~\cite{Bekenstein73a}
\begin{equation}
{\cal {\cal S}} = \frac{k_B c^3}{\hbar} \frac{A}{4G_N},
\label{Bek-Haw}
\end{equation}
where $A$ is the area of the horizon corresponding to the surface
at $r=r\Z{+}$. There are two important implications of this
result. Firstly, the entropy depends on both Planck's constant
$\hbar$ and Newton's constant $G_N$, hinting that black hole
thermodynamics unites quantum mechanics and gravity. Secondly, in
accordance with Wheeler's famous dictum, `black holes have no
hair', the formula (\ref{Bek-Haw}) is universal to all kinds of
black holes irrespectively of their charges, shapes and
rotation~\cite{Gibbons:1988,Sen:2007}.

Black hole solutions in curved backgrounds, especially, in an
anti-de Sitter (AdS) background, exhibit several interesting new
features which can be related to certain class of boundary field
theories residing in one dimension lower. Notably, the AdS
conformal field theory (CFT) (or gravity - gauge theory)
correspondence~\cite{Maldacena97a,Gubser:98a} has provided several
interesting insights into the dynamics of strongly coupled gauge
theories in the limit $N_c\to \infty$ and $\lambda =g\Z{\rm YM}^2
N_c \to \infty$, where $N_c$ is the number of colors (or the rank
of the gauge group) and $\lambda$ is the 't Hooft coupling.
Through AdS holography one can also study hydrodynamic properties
of a certain class of boundary CFTs. In particular, it has been
possible to compute, for a large class of four-dimensional CFTs
with an Einstein gravity dual, the ratio of the shear viscosity
$\eta$ to the entropy density $s$, which (in units where
$\hbar=k\Z{B}=1$) is given by~\footnote{The shear viscosity term
$\eta$ appears as a coefficient of frictional force in the spatial
traceless (dissipative) part of the stress-energy tensor
$T_{\mu\nu}(x)\equiv T_{\mu\nu}^{(0)}+ T_{\mu\nu}^{(1)}$, where
the first term describes the perfect (non-dissipative) fluid
dynamics, $T_{\mu\nu}^{(0)}=\left[\rho(x)+ p(x)\right] u\Z{\mu}(x)
u\Z{\nu}(x)-p(x) \delta_{\mu\nu}$ and $
T_{ij}^{(1)}=-\eta\left(\frac{1}{2}\nabla_i
u_j(x)+\frac{1}{2}\nabla_j u_i(x)-\frac{1}{3}\delta_{ij}\nabla
\cdot {\bf u}(x)\right)$, where the collective flow velocity
$u_\mu(x)\equiv (1, -{\bf u}(x))/\sqrt{1-{\bf
u}^2}$.}~\cite{Policastro:01,Kovtun03,Son-07},
\begin{equation}\label{eta-s}
\frac{\eta}{s} =\frac{1}{4\pi}\approx 0.08. \end{equation}  This
result has been conjectured to be a universal lower bound (the
so-called Kovtun-Starinets-Son (KSS) bound~\cite{KSS}) in
nature~\footnote{The KSS bound $\eta/s\ge 1/4\pi$ is perhaps
related to a similar limit of low energy absorption cross sections
for classical black holes~\cite{S.Das:1996}. Since $\eta\simeq
\frac{\sqrt{m k\Z{B} T}}{\sigma}$, a large (black hole) scattering
cross section $\sigma$ yields a small shear viscosity.}. While
this bound is below the value found for ordinary substances
(especially, cold atomic gases, including water and liquid
helium), a great deal of interest has been generated by results
from the Relativistic Heavy Ion Collider (RHIC) experiments
suggesting that just above the deconfinement phase transition (or
infrared limit), QCD is very close to saturating this bound (see,
e.g.,~\cite{Molnar05}).

For finite temperature quantum field theories, in particular, for
thermal super Yang-Mills theories at the weak coupling limit,
$g\Z{\rm YM}\ll 1$, both the shear viscosity and the entropy
density scale being proportional to $T^{3}$, namely $\eta\sim
\frac{\pi}{8} N_c^2 T^3$ and $s\sim \frac{\pi^2}{2} N_c^2 T^3$.
This scaling behaviour effectively looks similar to CFT one, for
which $\eta/s\sim 1/4\pi$~\cite{Policastro:01}. Notwithstanding
this result, supergravity calculations are generally carried out
in an opposite end, i.e., in the strong coupling regime where
$\lambda\equiv g\Z{YM}^2 N_c \gg 1$ and $N_c \to \infty$. This
limit is also known in the gauge theory as the 't Hooft limit,
where one takes the rank of the gauge group $N_c$ to infinity
while keeping $\lambda$ fixed. Presumably, without much surprise,
all gauge theories operate at a finite coupling. For example, QCD
is a gauge theory with $N_c=3$. Thus, to establish better contact
with QCD via AdS holography, it is essential to understand the
effect of $1/N_c$ corrections which arise from curvature square
corrections to Einstein-Hilbert action in the holographic
framework (see, for
example,~\cite{Henningson:98a,Narain99a,Ish02}).

Given that we do expect corrections to Einstein gravity to occur
in any consistent quantum theory of gravity, it is perhaps natural
to expect modifications to both the entropy law (\ref{Bek-Haw})
and the ratio $\eta/s=1/4\pi$. As the simplest modification to
classical Einstein gravity, the leading order higher derivative
corrections may be written as
\begin{equation}\label{main-action}
I\Z{\rm g}=\frac{1}{16\pi G_N}\int d^{d} x\sqrt{-g}
\left[R-2\Lambda +{\alpha^\prime} L^2 \left(a R^2+ b
R_{\mu\nu}R^{\mu\nu}+c
R_{\mu\nu\lambda\rho}R^{\mu\nu\lambda\rho}\right) \right],
\end{equation}
where $\alpha^\prime$ is a dimensionless coupling. For simplicity,
the bulk cosmological term $\Lambda$ is fixed in terms of the
length scale $L$, namely $\Lambda=-(d-1)(d-2)/2L^2$, which is a
sort of fine tuning! One may supplement the above action with
analogue Gibbons-Hawking surface terms~\cite{Myers:87}, but we do
not need these terms in our present discussion. In fact, the
Gauss-Bonnet (GB) term obtained by setting $a=c=1$ and $b=-4$ in
(\ref{main-action}) produces the most general action retaining
only second-order field equations and hence admits exact solutions
in numerous cases~\cite{Deser85a,HMaeda:08,Maeda-Naresh}. Since
the GB term is topological in $d=4$, especially, with a constant
coupling, $\alpha^\prime L^2=$ const, one considers a spactime for
which $d\ge 5$. The Gauss-Bonnet term is relevant not only because
of a solvability of the model but also due to its natural
appearance in lower dimensional actions of various compactified
string theories~\cite{Zwiebach:85a}. Moreover, it is a unique
combination of quadratic curvature invariants which is free of
ghost not only in a flat Minkowski background but also when
expanded about a warped AdS braneworld background~\cite{Ish02Y}.

We also note that the coefficient multiplying the GB term or terms
quadratic in curvature, has a mass dimension of $2$, or $({\rm
lenght})^2$; in (\ref{main-action}) we have replaced this
coefficient by $\alpha^\prime L^2$, so that $\alpha^\prime$ is
dimensionless. We focus most of our discussions below to the
gravity sector in AdS$_5$, for which we have (from AdS/CFT
dictionary~\cite{Maldacena97a,Narain99a})
$$ \frac{1}{16\pi G_N}\equiv \frac{N_c^2}{4\pi^2 L^3}.$$
Moreover, the curvature radius of AdS$_5$ can be defined by
$$ L=(4\pi g_s N_c)^{1/4} {\ell}_s,$$
with ${\ell}_s$ being the string scale and $g_s$ the string
coupling. In terms of the 't Hooft coupling $\lambda=g_{YM}^2
N_c$, the dimensionless scale $\alpha^\prime$ of string theory,
for instance, on AdS$_5 \times S^5$, is related to the SYM
parameters by $\alpha^\prime \sim 1/\sqrt{\lambda}$. The latter
implies that a small $\alpha^\prime$ corresponds to the strong
coupling limit, i.e. $g_{YM}^2 N_c \gg 1$ in dual supergravity
description. Our focus here is to compute the limits on black hole
entropy and shear viscosity nonperturbatively in the Gauss-Bonnet
coupling $\lambda_{GB}$ (or $\alpha^\prime\equiv \lambda_{GB}/2$),
so we hardly use the above relations, but they are useful for
expressing our results in terms of $N_c$ and the AdS curvature
$L$.

In fact, far from the classical limit ($N_c\to \infty$), or with
$\alpha^\prime L^2>0$, the ratio of shear viscosity to entropy in
eq.~(\ref{eta-s}) is expected to be modified. Accordingly, the KSS
bound $\eta/s\ge 1/(4\pi)$ could be violated at high energies or
short distances, or in the presence of generic higher derivative
corrections, $\alpha^\prime> 0$~\footnote{In gravitational
theories, the high energy effects arise through higher order
curvature effects.}. The reason is simple to understand. In
contrast to entropy density $s$ which is the zeroth order
parameter in ideal fluid equations, the first order transport
coefficients in macroscopic hydrodynamic equations for
non-equilibrium systems, such as shear viscosity $\eta$ and heat
conductivity $\kappa$ can take a nearly vanishing value at high
densities and temperatures. The ratio $\eta/s$ is perhaps not
limited by any quantum bound.

In ref.~\cite{Molnar:08b}, it has been argued that in the very
center of the collision zone at RHIC, $\eta/s$ may take a value as
small as $\eta/s \approx 0.4/(4\pi)$, which is less than the
conjectured KSS bound. Further, in~\cite{Brigante-etal} it has
been shown that, for a class of CFTs in flat space ($\epsilon=0$)
with Gauss-Bonnet gravity dual, the ratio $\eta/s$ is given by
\begin{equation}
\frac{\eta}{s}=\frac{1}{4\pi}\left(1-\frac{2(d-1)\lambda\Z{GB}}{(d-3)}\right),
\end{equation}
where $\lambda\Z{\rm GB} \equiv (d-3)(d-4)\alpha^\prime$, which is
smaller than $1/4\pi$ for $\lambda\Z{\rm GB}>0$. In the $d=5$
case, $\lambda\Z{\rm GB}< 1/4$ is required to keep $\eta/s$
positive definite. In~\cite{Brigante-etal}, based on bulk causal
structures of an AdS$_5$ black brane solution, a more stronger
bound for $\lambda\Z{\rm GB}$ was proposed, namely
\begin{equation}
\lambda\Z{\rm GB} < \frac{9}{100}
\end{equation}
or equivalently
\begin{equation} \frac{\eta}{s}\ge \frac{16}{25}
\left(\frac{1}{4\pi}\right),
\end{equation}
which otherwise violates a microcausality in the dual CFT defined
on a flat hypersurface. In~\cite{Brigante-etal}, in the context of
Gauss-Bonnet gravity, it was found hard to understand the sudden
change in behaviour at $\lambda=9/100$. In this paper we provide a
plausible explanation for this change in behaviour. We actually
establish that the critical value of $\lambda\Z{\rm GB}$ beyond
which the theory becomes inconsistent is related to the entropy
bound for a class of AdS GB black holes. We also find that, in the
holographic context, the AdS-GB black hole solutions with
spherical and hyperbolic event horizons ($\epsilon=\pm 1$) allow
much wider possibilities for $\eta/s$.

\section{Gauss-Bonnet Gravity and Causality Violation}

Let us start with a pure AdS-GB black hole solution (i.e. without
any electric or magnetic charge~\footnote{Classical properties of
Gauss-Bonnet black holes with Maxwell type electric or magnetic
charges will be briefly discussed in the appendix.}), for which
the entropy and Hawking temperature are given by~\cite{Ish02}
\begin{equation}\label{GB-entropy}
{\cal {\cal S}} = \frac{A}{4G_N}\left(1+ \frac{2(d-2)\epsilon
\lambda\Z{\rm GB} }{(d-4)x^2} \right),\quad
 T=\frac{(d-1)x^4+ \epsilon(d-3)x^2+ (d-5)\epsilon^2 \lambda\Z{\rm GB}}
 {4\pi L x(x^2+2 \epsilon \lambda\Z{\rm
GB})}
\end{equation}
(in units $c=\hbar=k_B=1$) where $x\equiv {r\Z{+}}/{L}$, $A \equiv
V\Z{d-2} r\Z{+}^{d-2}$, with $V\Z{d-2}$ being the unit volume of
the base manifold or the hypersurface ${\cal M}$. The above
entropy relation holds with an arbitrary $\Lambda$, i.e. even if
$\Lambda=0$.

\begin{figure}[ht]
\begin{center}
\hskip-0.3cm
\epsfig{figure=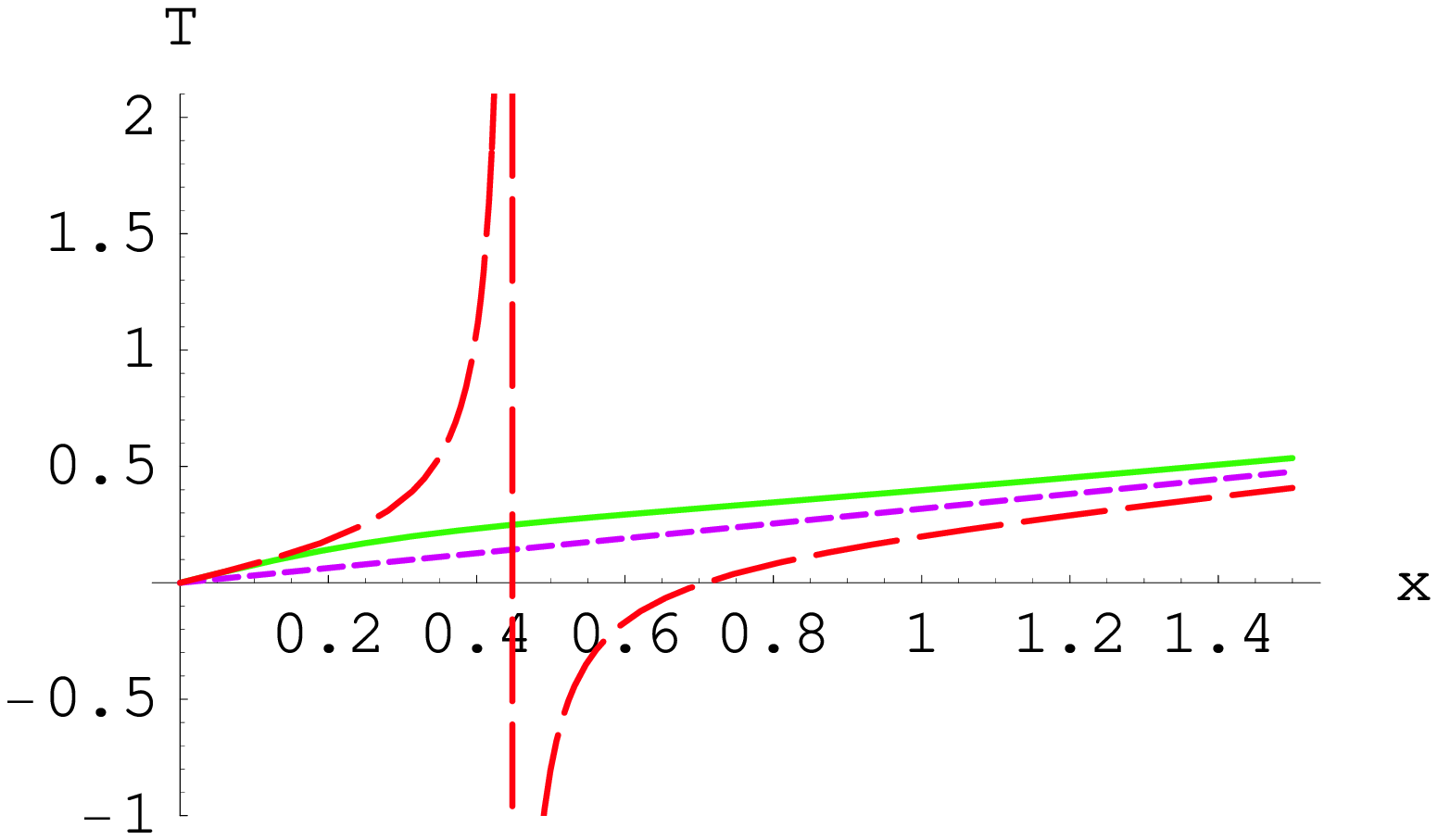,height=2.1in,width=2.9in}
\hskip0.2cm
\epsfig{figure=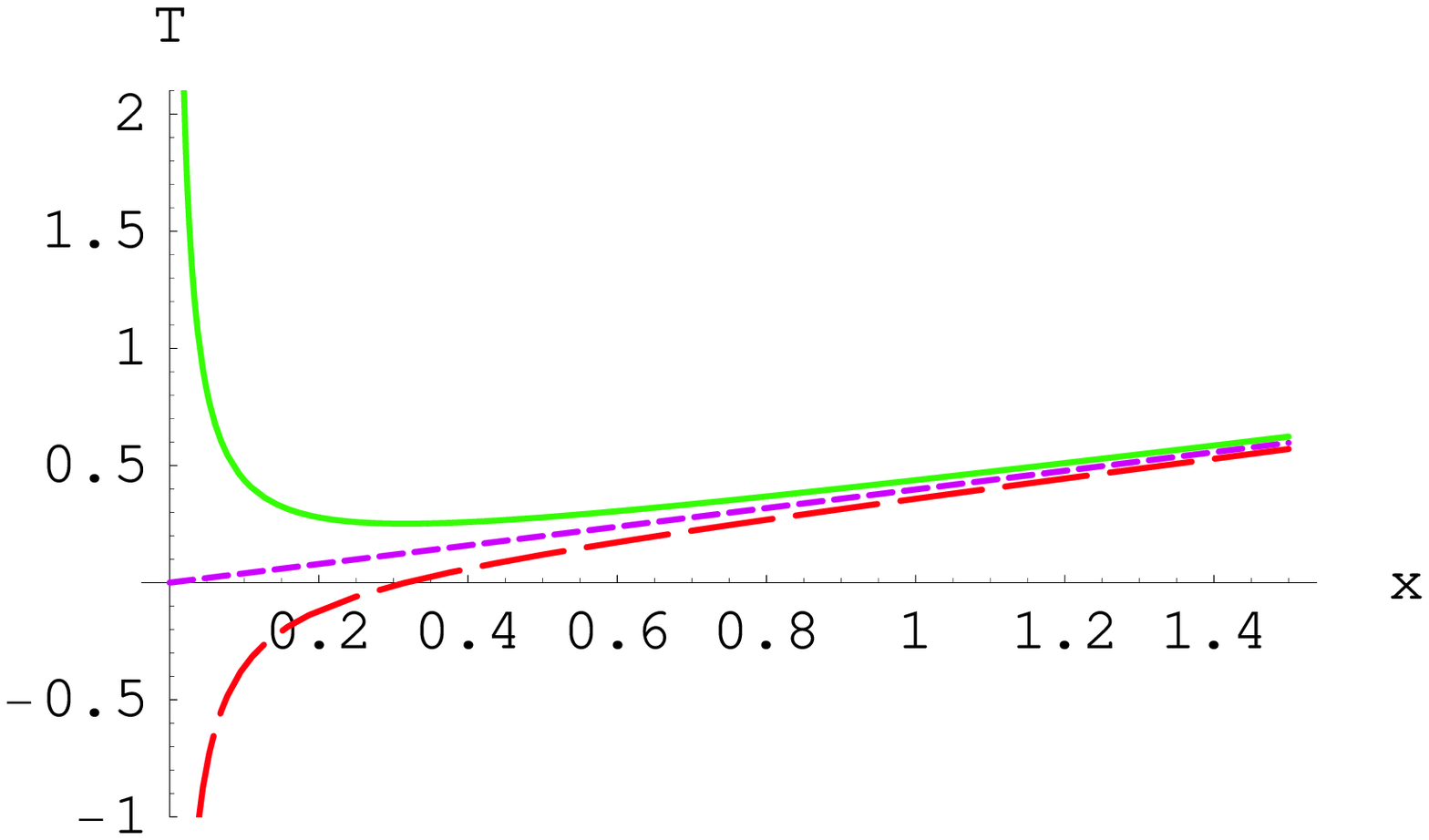,height=2.1in,width=2.9in}
\end{center}
\caption{\label{fig1} The Hawking temperature as a function of $x$
(left plot) with $d=5$, $\lambda\Z{\rm GB}=0.1$ and (right plot)
with $d=6$, $\lambda_{\rm GB}=0.25$. The solid (green), short-dash
(purple) and long-dash (red) lines correspond, respectively, to
$\epsilon=+1$, $\epsilon=0$ and $\epsilon=-1$.}
\end{figure}

Notice that, unlike in Einstein's general relativity, the entropy
of a GB black hole depends on the curvature constant $\epsilon$,
whose value determines the geometry of event horizon
\begin{equation}
{\cal M} =\left\{ \begin{array}{ll} {\rm S}^{d-2} : {\rm
Euclidean\; de\; Sitter\; space\;}
~~ (\epsilon=+1)\\
{\rm I\!\, \! R}^{d-2} : {\rm Ricci\; flat\; space\;} ~~(\epsilon=0)\\
{\rm H}^{d-2} : {\rm Euclidean\; Anti\; de\; Sitter\; space\;}
(\epsilon=-1).
\end{array} \right.
\end{equation}
The hypersurface ${\cal M}$ is related to the ${\cal M}^\prime$ on
which the dual field theory is defined only by a rescaling of the
metric. As a consequence, the Hawking temperature of a boundary
conformal field theory can be different by some constant, say
$N_*$, which specifies the speed of light of the boundary theory.
In the particular case where $\epsilon=0$, the Hawking temperature
is found to be proportional to black hole horizon size, namely
\begin{equation}
T\Z{\rm CFT}=N_* \frac{(d-1) r_+}{4\pi L^2}.
\end{equation}
Particularly, on a flat hypersurface at $r= \infty$, we have (see
eq.~(\ref{sol-N*}) below)
\begin{equation}
f(r) \to \frac{r^2}{a^2\, L^2}, \quad {\rm with}\quad {a^2} =
\frac{1}{2}\left(1+\sqrt{1-4\lambda_{\rm GB}}\right).
\end{equation}
If we choose $N_\ast \equiv a$, then the boundary speed of light
is unity. In the $\epsilon=0$ case, we then find $s\equiv {\cal
S}/V \sim \frac{1}{4G_N} (r_+)^{d-2} \sim \frac{\pi^2}{2} N_c^2
T^{d-2}$. However, such a scaling relation does not hold, in
general, with $\alpha^\prime$ corrections, especially, when
$\epsilon= \pm 1$. This behavior can be seen also from the plots
in figure~\ref{fig1}. To quantify this, let us consider the
relation between $r_+$ and $T$. In the $d=5$ case, we find
\begin{equation}
r_+|\Z{T\to 0}=4\pi L^2 \lambda\Z{\rm GB} T + {\cal O}(T^2), \quad
r_+|\Z{T\to \infty} = \pi L^2 T + {\cal O} (T^{-1}).
\end{equation}
The Hawking temperature roughly scales in between $\frac{r_+}{\pi
L^2} \frac{1}{4\lambda\Z{\rm GB}}$ and $\frac{r_+}{\pi L^2}$. When
$\epsilon=+1$, the entropy density is $s={\cal
S}/V=\frac{1}{4G_N}\left(r_+^3+6 \lambda\Z{\rm GB} r_+
L^2\right)$, which increases with $T$ more rapidly (as compared to
that in Einstein gravity). It is quite plausible that the ratio
$\eta/s$ becomes less than $1/4\pi$ in the regime where the
Gauss-Bonnet contribution to entropy exceeds (or becomes
comparable) to that of Einstein-Hilbert term.

\begin{figure}[ht]
\begin{center}
\hskip-0.3cm
\epsfig{figure=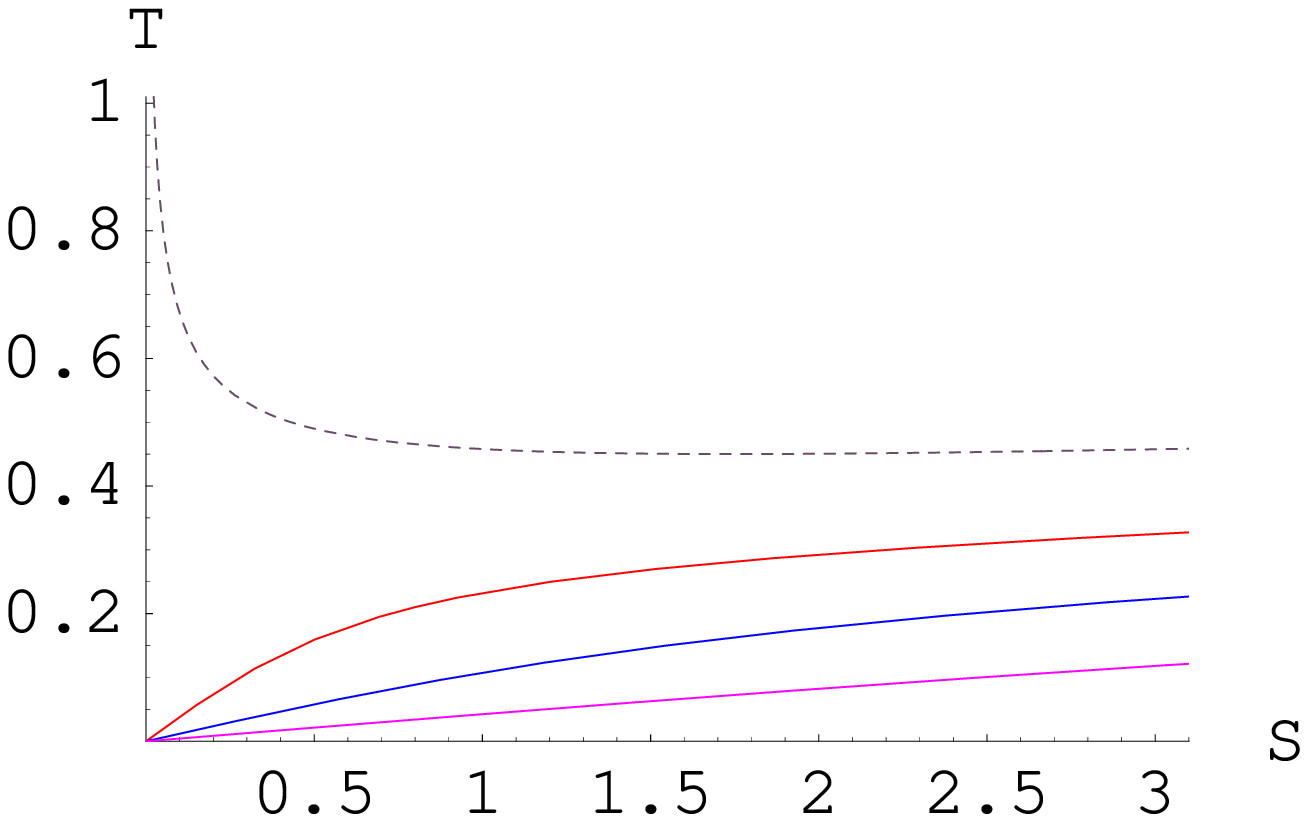,height=2.1in,width=2.9in}
\hskip0.2cm
\epsfig{figure=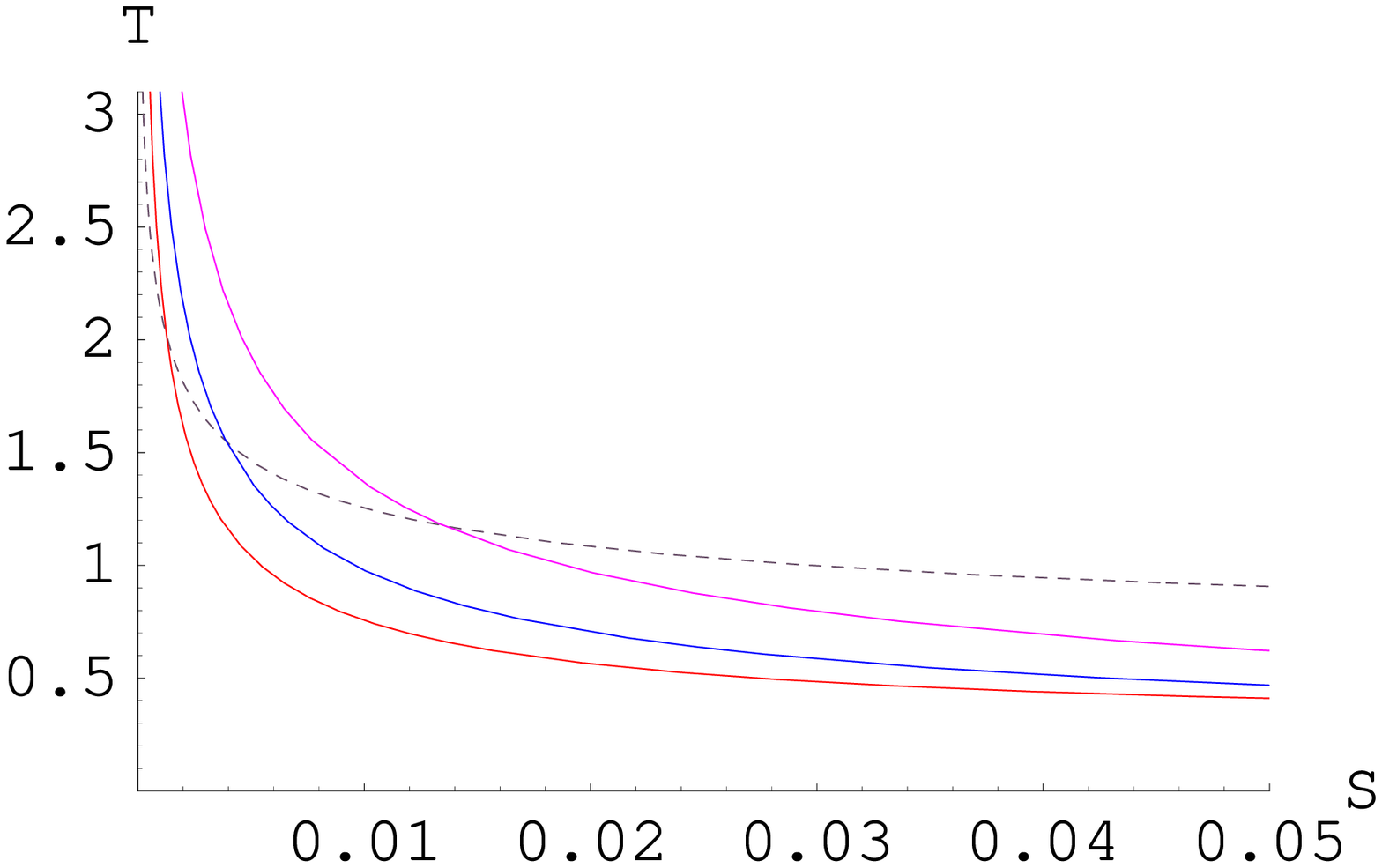,height=2.1in,width=2.9in}
\end{center}
\caption{\label{fig2} The Hawking temperature vs entropy with
$\epsilon=+1$ (spherical black hole). (Left plot) $d=5$,
$\lambda_{\rm GB}=1/12, 0.15, 0.25$ (top to bottom) and (right
plot) $d=6$, $\lambda_{\rm GB}= 0.135, 0.25, 0.5$ (top to bottom).
In each plot the dotted line corresponds to $\lambda\Z{\rm GB}=0$.
}
\end{figure}
\begin{figure}[ht]
\begin{center}
\hskip-0.3cm
\epsfig{figure=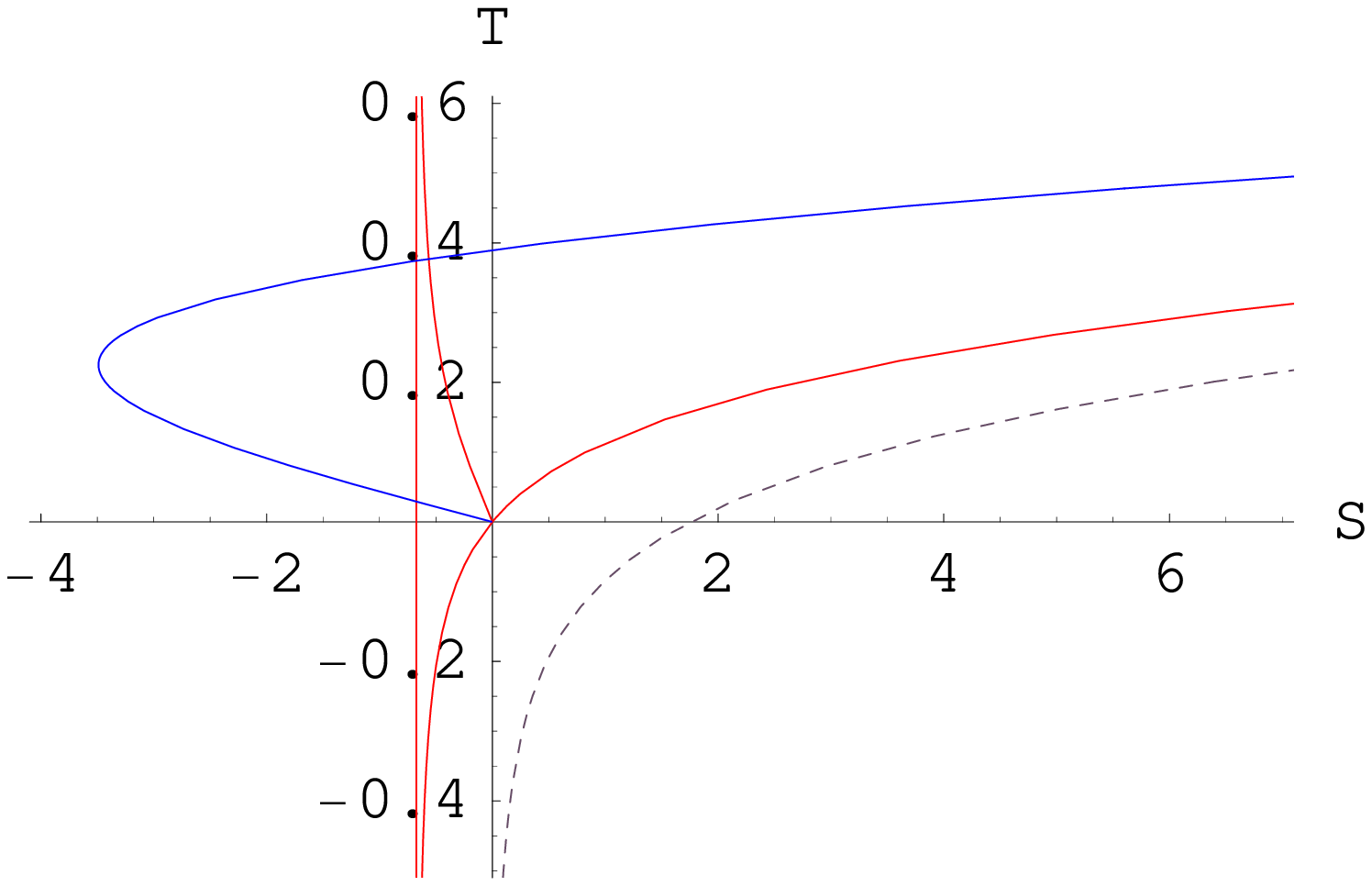,height=2.1in,width=2.9in}
\hskip0.2cm
\epsfig{figure=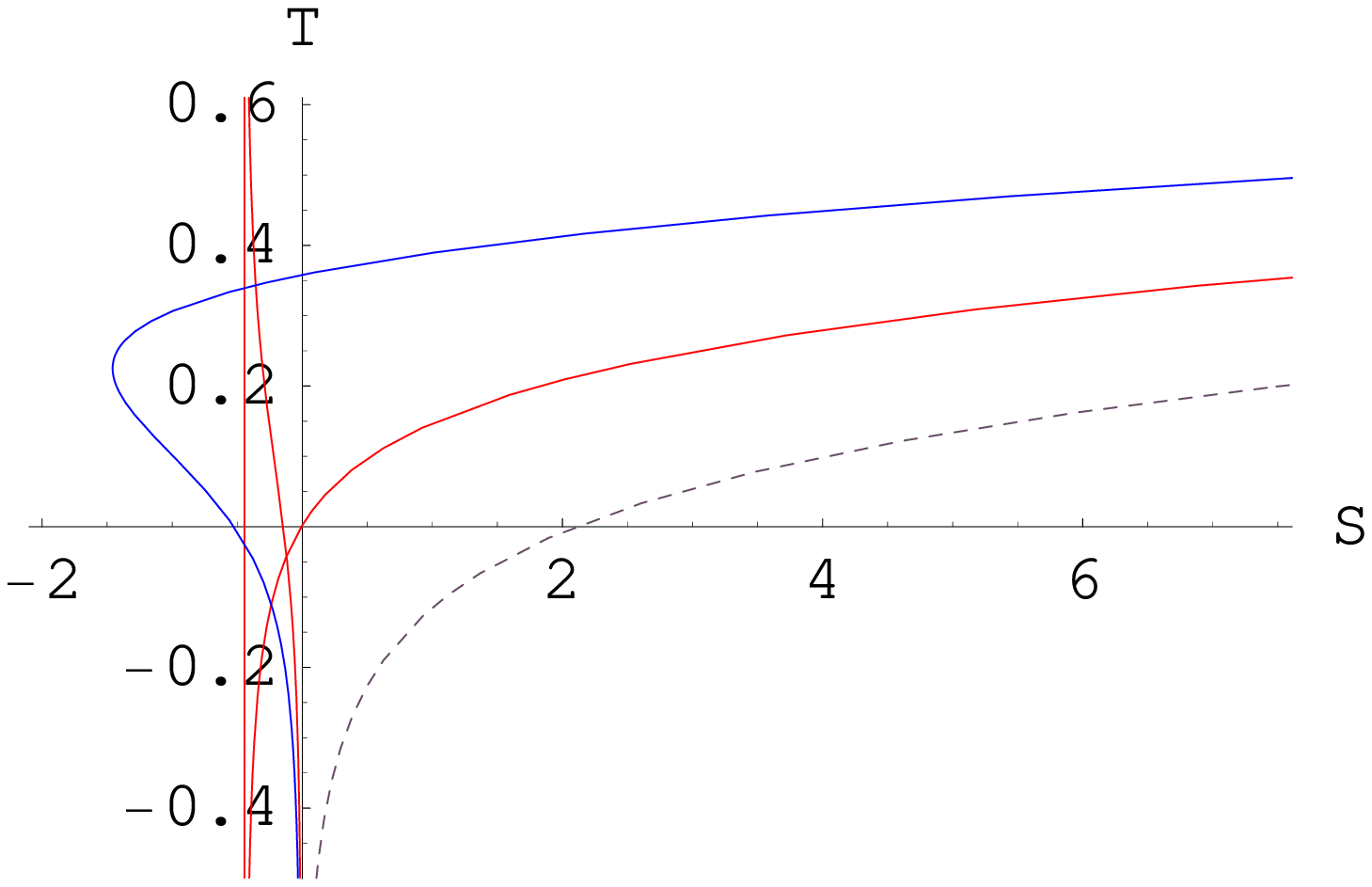,height=2.2in,width=2.9in}
\end{center}
\caption{\label{fig3} The Hawking temperature vs entropy with
$\epsilon=-1$. (Left plot) $d=5$, $\lambda_{\rm GB}=0, 0.0834,
0.25$ and (right plot) $d=6$, $\lambda_{\rm GB}=0, 0.138, 0.25$
(top to bottom).}
\end{figure}

We also note that, for $\epsilon=+1$, the Hawking temperature of
an AdS$_5$ GB black hole is always less than that of a
Schwarzschild black hole, for a given (fixed) entropy. Also, the
entropy vanishes as $T\to 0$. In the $d=6$ case, the Hawking
temperature not only exceeds that of a Schwarzschild black hole,
at a small $r_+$, but it also diverges as ${\cal S}\to 0$ (see
figure~\ref{fig2}). From this observation, we can argue that the
$d=5$ case is the most relevant one.

A particularly interesting case is $\epsilon=-1$, for which the
Hawking temperature of an AdS-GB black hole diverges to $-\infty$
as $x\to 0$. But in this limit the black hole entropy also
diverges to $-\infty$. Thus, in order to properly understand the
behaviour of temperature at a small value of $r_+/L$, which
corresponds to a large $z\equiv r/r\Z{+}$ or infrared limit of a
dual CFT, one should actually study the behaviour of temperature
as a function of entropy. The plots in figure \ref{fig3} show
that, for $\epsilon=-1$, the entropy ${\cal S}$ can have an
extremum as a function of Hawking temperature. In this case,
beyond a critical coupling $\lambda\Z{\rm GB}
> \lambda\Z{\rm crit}$ the entropy ${\cal S}$ becomes
negative at zero Hawking temperature, indicating a violation of
cosmic censorship or the second law of the thermodynamics. In the
AdS$_5$ case, we find that $\lambda\Z{\rm crit}=1/12$. It is {\it
not} a coincident that this critical value of $\lambda\Z{\rm GB}$
above which the theory is inconsistent nearly coincides with the
bound $\lambda\Z{\rm GB}< 0.09$ required for a consistent
formulation of a class of CFTs in flat space with Gauss-Bonnet
gravity dual.

\subsection{Shear Viscosity for Gauss-Bonnet Gravity}

The effect of Gauss-Bonnet coupling $\lambda\Z{\rm GB}$ on shear
viscosity $\eta$ can be studied by considering small metric
fluctuations $\phi=h^1_2$ around an AdS black hole solution of the
form
\begin{equation}
ds^2=-f(r) N\Z{\ast}^2 dt^2 + \frac{1}{f(r)} dr^2 +
\frac{r^2}{L^2} \left(\frac{d x\Z{3}^2}{1-\epsilon x\Z{3}^2}+
x\Z{3}^2 \sum_{i=1}^{2} dx_i^2 + 2\phi(t, x\Z{3} ,r) dx_1
dx_2\right)
\end{equation}
where $\epsilon=0, \pm 1$ and
\begin{equation}\label{sol-fr}
f(r)=\epsilon+\frac{r^2}{L^2} \frac{1}{2\lambda\Z{\rm GB}}
\left[1\pm \sqrt{1-4\lambda\Z{\rm GB}+\frac{4\lambda\Z{\rm GB}
r\Z{+}^4}{r^4}\left(1+\frac{\epsilon
L^2}{r\Z{+}^2}+\frac{\lambda\Z{\rm GB}\epsilon^2
L^4}{r\Z{+}^4}\right)}\right].
\end{equation}
Although there are two distinct vacuum solutions, we shall
consider only the negative root in eq.~(\ref{sol-fr}) which has a
smooth limit to Einstein gravity, $\lambda\Z{\rm GB}\to
0$~\footnote{For an electrically charged AdS Gauss-Bonnet black
hole, these two branches of solutions may actually join together
without developing a physical singularity (see the appendix).}.
Note that as $r\to \infty$,
\begin{equation}\label{sol-N*}
f(r)\to \frac{r^2}{\tilde{a}^2 L^2}, \qquad \frac{1}{\tilde{a}^2}=
\left[\frac{\tilde{\epsilon}}{z^2} + \frac{2} {1+
\sqrt{1-4\lambda\Z{\rm GB}}}\right],
\end{equation}
where $\tilde{\epsilon}\equiv \epsilon/x^2$, $x\equiv r\Z{+}/L$
and $z\equiv r/r\Z{+}$. To allow a black hole interpretation, the
GB coupling must satisfy $\lambda\Z{\rm GB} \le 1/4$ (see,
e.g.,~\cite{Wiltshire:88u,Ish02}); beyond this value, the above
solution does not admit a consistent vacuum AdS solution (see the
appendix).

In the $\epsilon=  \pm 1$ case, the curvature of the boundary or
the hypersurface ${\rm S}^3$ (or ${\rm H}^3$) introduces a new
scale into the problem. However, the result in flat space is
reproduced as a limit, which can be characterized as the high
temperature limit ($z \to \infty$). To be specific, we make the
following ansatz
\begin{equation}
\phi(t, x\Z{3},r)=\int \frac{dw dq}{(2\pi)^3} \, \phi(r ; k)\,
e^{-iwt+iq x_3}, \quad \phi(r; -k)=\phi^*(r,k)
\end{equation}
(where $k=(w, 0, 0,q)$). The quadratic action for $\phi$ takes the
form (up to the surface terms)
\begin{equation}\label{quad-action}
\delta^2 I\propto \int dz\, \frac{dw\, dq}{(2\pi)^2} \left(K
(\partial_z\phi)^2-K_2\phi^2\right),
\end{equation}
where
\begin{equation}
K=z^2 \tilde{f} \left(z-{\lambda_{\rm
GB}\partial_z\tilde{f}}\right), \quad K_2=\frac{z^2 \tilde{w}^2}{
N_*^2 \tilde{f}}\left(z-{\lambda_{\rm
GB}\partial_z\tilde{f}}\right) -z (1-\lambda\Z{\rm GB}
\partial_z^2
\tilde{f})\left(\tilde{q}^2+2\tilde{\epsilon}\right)
\end{equation}
and
\begin{equation}
\tilde{f}(z)=\frac{L^2}{r\Z{+}^2}\,f(r)=\tilde{\epsilon} +
\frac{z^2}{2\lambda\Z{\rm GB}} \left(1- \sqrt{1-4\lambda\Z{\rm
GB}+\frac{4\lambda\Z{\rm GB}}{z^4}
\left(1+\tilde{\epsilon}+\lambda\Z{\rm GB}
\tilde{\epsilon}^2\right)}\right),
\end{equation}
where $\tilde{\epsilon}=\epsilon/x^2$, $x\equiv r\Z{+}/L$,
$z\equiv r/r\Z{+}$, $\tilde{w}\equiv w L/x$, $\tilde{q}\equiv q
L/x$ and $\tilde{f}\equiv f/x^2$. The equation of motion following
from (\ref{quad-action}) is~\footnote{Especially, in $d=5$, the
linearised equation of motion can be expressed in this simple
form. In dimensions $d\ge 6$, various functions ($K$, $K_2$)
receive extra contributions being proportional to
$(\tilde{\epsilon}-\tilde{f})/(z
\partial_z\tilde{f})$, see also~\cite{Dotti:04a} for a
discussion on gravitational instability of six-dimensional
asymptotically flat EGB black holes.}
\begin{equation}\label{main-eqn-phi} K
\partial_z^2 \phi +
\partial_z K \partial_z\phi + K_2\phi =0,
\end{equation}
Especially, for $\epsilon=0$, in the limit $z\to 1$, we
find~\footnote{At this point, one also notes the relation $
\partial_z\tilde{f}=[4z^3+2z(\tilde{\epsilon}-\tilde{f})]/
[z^2+2\lambda\Z{\rm GB}(\tilde{\epsilon}-\tilde{f})].$}
\begin{eqnarray}
\frac{K_2}{K}&=&\frac{\tilde{w}^2}{16N_*^2 (z-1)^2}+
\left(\frac{\tilde{w}^2(1-8\lambda\Z{\rm GB})}{16
N_*^2}-\frac{\tilde{q}^2+2\tilde{\epsilon}}{4}\right)\frac{1}{z-1}+{\cal
O}(1),
\nonumber\\
\frac{\partial_zK}{K}&=&\frac{1}{z-1}+\frac{5+24\lambda\Z{\rm
GB}}{2} +{\cal O}(z-1).
\end{eqnarray}
For $\epsilon\ne 0$, the limiting values of these functions are
regular at $z=1$, but their explicit expressions will not be
important in the present discussion.

From the viewpoint of gravity--gauge theory duality, it is
instructive to study the large $z$ behaviour of AdS solutions. In
the limit $z\to \infty$, we find
\begin{eqnarray}\label{K-K2-largez}
\frac{K_2}{K}&=& \frac{\tilde{w}^2\lambda\Z{\rm GB}^2}{N_*^2 a^4
z^4}- \frac{\lambda\Z{\rm GB}(\tilde{q}^2+2\tilde{\epsilon})}{a^2
z^4}
+{\cal O}(z^{-6}),\nonumber \\
\frac{\partial_z K}{K} &=& \frac{5}{z}-\frac{2\lambda
\tilde{\epsilon}}{a^2 z^3}+{\cal O} (z^{-5}),
\end{eqnarray}
where $a^2 \equiv \frac{1}{2} (1+ \sqrt{1-4\lambda\Z{\rm GB}})$.
The effect of the curvature $\tilde{\epsilon}$ on a dual field
theory can appear only as a small correction to the results in
flat space~\cite{Brigante-etal}. To be precise, one solves
eq.~(\ref{main-eqn-phi}) with the boundary condition
\begin{equation}
\phi(z; k)= a\Z{\rm in}(k) \phi\Z{\rm in}(z;k)+ a\Z{\rm out}(k)
\phi\Z{\rm out}(z;k),
\end{equation}
$a\Z{\rm out}\equiv 0$ and $a\Z{\rm in}\equiv J(k)$, where $J(k)$
is an infinitesimal boundary source for the fluctuating field
$\phi$. Following~\cite{Brigante-etal}, we make the following
low-frequency expansion
\begin{equation}\label{ansatz-small-w} \phi\Z{\rm in} (z; w, q)=
\exp\left[-\frac{i\tilde{w}}{4N_*} \ln \left(\frac{\tilde{a}^2
\tilde{f}}{z^2}\right)\right]\times
\left(1-i\frac{\tilde{w}}{4N_*} g_1(z)+{\cal O}(\tilde{w}^2,
\tilde{q}^2)\right).
\end{equation}
Note that as $z\to \infty$,
\begin{eqnarray}
\frac{\tilde{a}^2 \tilde{f}}{z^2}&\to&
1-\frac{C}{\sqrt{1-4\lambda\Z{\rm GB}}}\frac{\tilde{a}^2}{z^4}+
{\cal O} (1/z^8),\quad {\rm where} \quad \tilde{a}^2\equiv
\left(\frac{\tilde{\epsilon}}{z^2}+\frac{1}{a^2}\right)^{-1}
\nonumber \\
\ln\left(\frac{\tilde{a}^2\tilde{f}}{z^2}\right) &\to &
 -\frac{C
a^2}{\sqrt{1-4\lambda\Z{\rm GB}}\, z^4} +\frac{C\tilde{\epsilon}
a^2}{\sqrt{1-4\lambda\Z{\rm GB}}\, z^6}+ {\cal
O}(1/z^8),\label{expand-expo}
\end{eqnarray}
where $C\equiv 1+\tilde{\epsilon} +\lambda\Z{\rm
GB}\tilde{\epsilon}^2$. To find the large $z$ behavior of
$g_1(z)$, one effectively solves the equation
$K\phi^{\prime\prime}+\partial_z K\phi^\prime= 0$, which yields
\begin{equation}
 g_1(z)=\frac{(C-1+4\lambda\Z{\rm GB})}{\sqrt{1-4\lambda\Z{\rm
 GB}}}\frac{a^2}{z^4}-\frac{C\tilde{\epsilon}}{\sqrt{1-4\lambda\Z{\rm
 GB}}}\frac{a^2}{z^6}
 +\frac{2\lambda\Z{\rm GB}\tilde{\epsilon}\sqrt{1-4\lambda\Z{\rm GB}}}{3z^6} +{\cal
O}(z^{-8}).
\end{equation}
Substituting this value of $g_1(z)$, along with
eq.~(\ref{expand-expo}), into eq.~(\ref{ansatz-small-w}), we can
see that the effect of curvature on $\phi(z; k)$ is only
sub-leading in the limit $z\to \infty$. More precisely,
\begin{equation} \phi(z; k) =J(k) \left[1+\frac{i\tilde{w}}{4N_*}
a^2 \sqrt{1-4\lambda\Z{\rm
GB}}\left(\frac{1}{z^4}-\frac{4\lambda\Z{\rm GB}
\tilde{\epsilon}}{3(1+\sqrt{1-4\lambda\Z{\rm GB}})}\frac{1}{z^6}+
{\cal O}(z^{-8})\right)+{\cal O}(\tilde{w}^2 \right].
\end{equation}
From this we can see that the curvature on a boundary does not
affect the shear viscosity
\begin{equation}\label{main-result}
{\eta}=\frac{1}{16\pi G_N} \left(\frac{r\Z{+}^3}{L^3}\right)
(1-4\lambda\Z{\rm GB})
\end{equation}
obtained by the Kubo formula
\begin{equation}\label{Kubo-formula}
\eta = \lim_{w\to 0}
\frac{1}{2iw}\left[G^A_{12,12}(w,0)-G^R_{12,12}(w,0)\right] \equiv
\lim_{w\to 0} \frac{1}{w} {\rm Im} G^R_{12, 12} (w,0),
\end{equation}
which relates $\eta$ to zero spatial momentum ($q=0$), low
frequency limit of the retarded two-point Green's
function~\footnote{AdS/CFT relates every field in supergravity (or
gravity solutions in AdS spaces) to a corresponding gauge
invariant operator of dual gauge theory, such as SYM
theory~\cite{Gubser:98a}. In particular, the two-point correlation
function of $T\Z{1 2}$ corresponds to scalar fluctuations $\phi^i$
in gravity, whose boundary values at infinity ($z\to \infty$) are
$\phi_\infty^i$. The CFT partition function corresponds to the AdS
action, $\langle \exp \left(\int \phi_\infty^i {\cal O}_i\right)
\equiv \exp \left(-I_{AdS} [\phi^i]\right)$. For EGB gravity, the
on-shell action (which utilizes the equations of motion) reduces
to surface contribution: $I[\phi^\infty (z)]=-\frac{r_+^4
N_*}{32\pi G_5 L^5}\int \frac{dw dq}{(2\pi)^2}\left(K
\partial_z\phi(z) \phi(z)+\cdots \right)|\Z{\rm surface}$
(see~\cite{Brigante-etal} for other details).}.
\begin{equation}
G^{R}_{12,12}(w,0)= -i \int d^4 x e^{iwt}\theta(t) \langle
[T_{12}(t,{\vec x}),T_{12}(0,{\vec 0})]\rangle =i\eta w + {\cal
O}(w^2)
\end{equation}
(modulo contact terms) and $G^A(w, \vec{q})\equiv
G^R(w,\vec{q})^*$. One also notes that, in the limit $\lambda_{\rm
GB}\to 1/4$, the retarded Green's function in AdS spacetimes do
not receive any corrections from the massive Kaluza-Klein modes.
This result is consistent with the discussion in~\cite{Ish02Y}, in
reference to Randall-Sundrum type warped braneworld models.

Finally, the ratio $\eta/s$ is given by~\footnote{A decade ago,
studies by Gubser-Klebanov-Polyakov/Witten~\cite{Gubser:98a}
showed that AdS/CFT works well for spherically symmetric AdS black
holes. Nevertheless, the Kubo formula could appear somewhat formal
on $S^3$, the reason being that $S^3$ is a finite-size manifold.
In turn, one could ask whether there exists a proper hydrodynamic
limit for a class of CFTs defined on $S^3$. In that respect,
hyperbolic black holes may introduce new and fruitful feature.
Here we only assume that the Kubo formula is applicable to a
slightly negatively curved spacetime, namely on ${\cal H}^3$, but
it would be really nice to check this point. Our expectation seems
reasonable from the viewpoints that hyperbolic black holes are
usually described as thermal Rindler states of the dual conformal
field theory in flat space and the Kubo formula is a good
approximation in the AdS asymptotic ($z\to \infty$) where the
effect of curvature becomes negligibly small.}
\begin{equation}
\frac{\eta}{s} =\frac{1}{4\pi} \frac{(1-4\lambda\Z{\rm
GB})}{(1+6\tilde{\epsilon} \lambda\Z{\rm GB})}.
\end{equation}
Given that $0<\lambda\Z{\rm GB}<1/4$, for a hyperbolic AdS-GB
black hole, the ratio $\eta/s$ can be larger than $1/4\pi$, while
for a spherical AdS-GB black hole, $\eta/s$ may further be
decreased (as compared to the $\epsilon=0$
case~\cite{Brigante-etal,Dutta:2008gf}). It seems possible to
saturate the KSS bound $\eta/s\ge 1/4\pi$ in the $\epsilon=-1$
case; specifically, at some fixed value of $\tilde{\epsilon}$,
namely $\epsilon=-1$ and $x=\sqrt{3/2}$, one finds
$\eta/s=\frac{1}{4\pi}$. Of course, with $\epsilon=0$, it also is
possible to obtain $\eta/s > 1/4\pi$ by taking $\lambda\Z{\rm
GB}<0$, but in this case the background solution may contain
ghosts.

Although the result (\ref{main-result}) is obtained in a high
temperature limit, let us assume that in a thought experiment one
can minimise the entropy of a given state. The minimum of entropy
density occurs for the $\epsilon=-1$ extremal solution at
$x=\sqrt{1/2}$~\footnote{In $d=5$, the location of the extremal
horizon of an AdS GB black hole doe not depend on the strength of
the coupling $\lambda\Z{\rm GB}$~\cite{Ish04C}.}, which is given
by
\begin{equation}\label{GB-extremal-s}
s=\frac{1}{G_N} \frac{1}{2^{7/2}}\left(1-12\lambda\Z{\rm
GB}\right).
\end{equation}
At this extremal state the shear viscosity is given by
\begin{equation}
\eta=\frac{1}{4\pi G_N} \frac{1}{2^{7/2}}\left(1-4\lambda\Z{\rm
GB}\right)
\end{equation}
Hence, with $\lambda\Z{\rm GB}\le 1/12$ (as implied by the
positivity of extremal entropy), we find~\footnote{In the
$\epsilon=+1$ case, the lower bound on $\eta/s$ could be very
close to that in flat space which however arises as a consequence
of boundary causality.}
\begin{equation}\label{limits-eta}
\frac{\eta}{s}\ge \frac{2}{3} \left(\frac{1}{4\pi}\right) \quad
(\epsilon=0), \qquad  \frac{\eta}{s} \le \frac{5}{3}
\left(\frac{1}{4\pi}\right) \quad (\epsilon=-1).
\end{equation}
Remarkably, the lower bound $\eta/s\approx 0.66/4\pi$ is similar
to a lower value of $\eta/s$ anticipated in the very center of
collision at RHIC (see, e.g.~\cite{Molnar:08b}).

\begin{figure}[ht]
\begin{center}
\hskip-0.3cm
\epsfig{figure=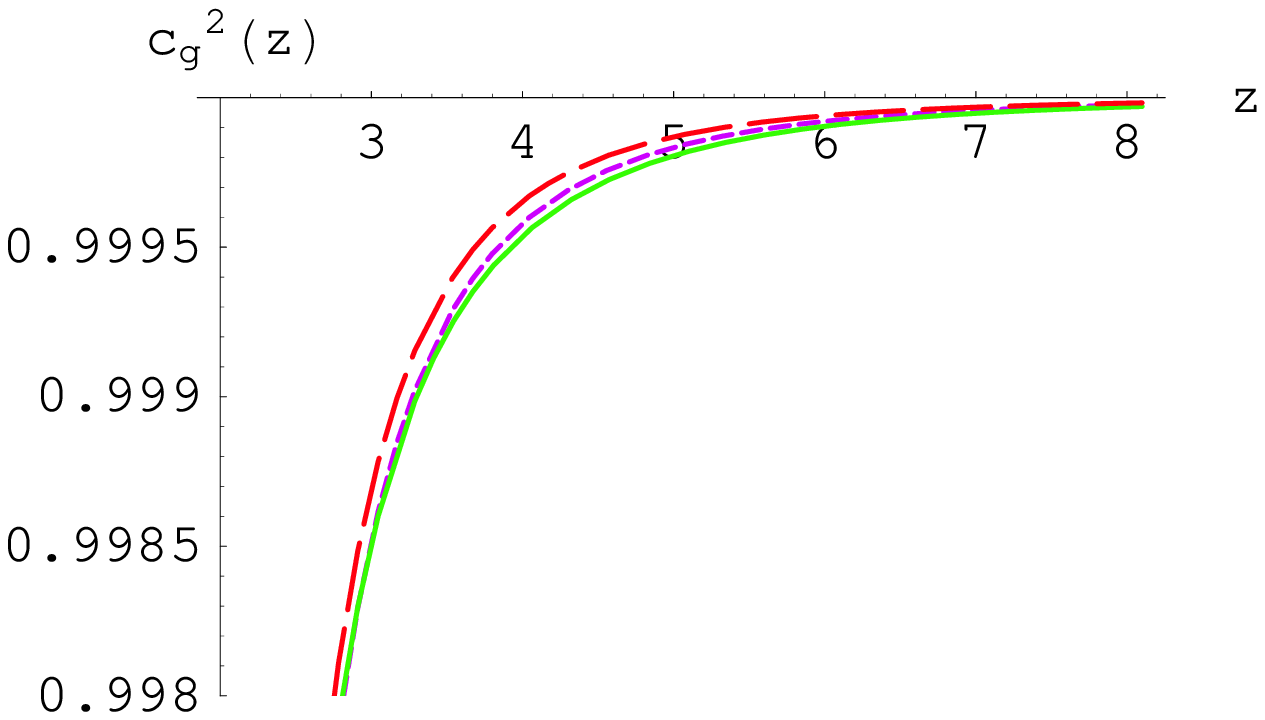,height=2.1in,width=2.9in}
\hskip0.2cm
\epsfig{figure=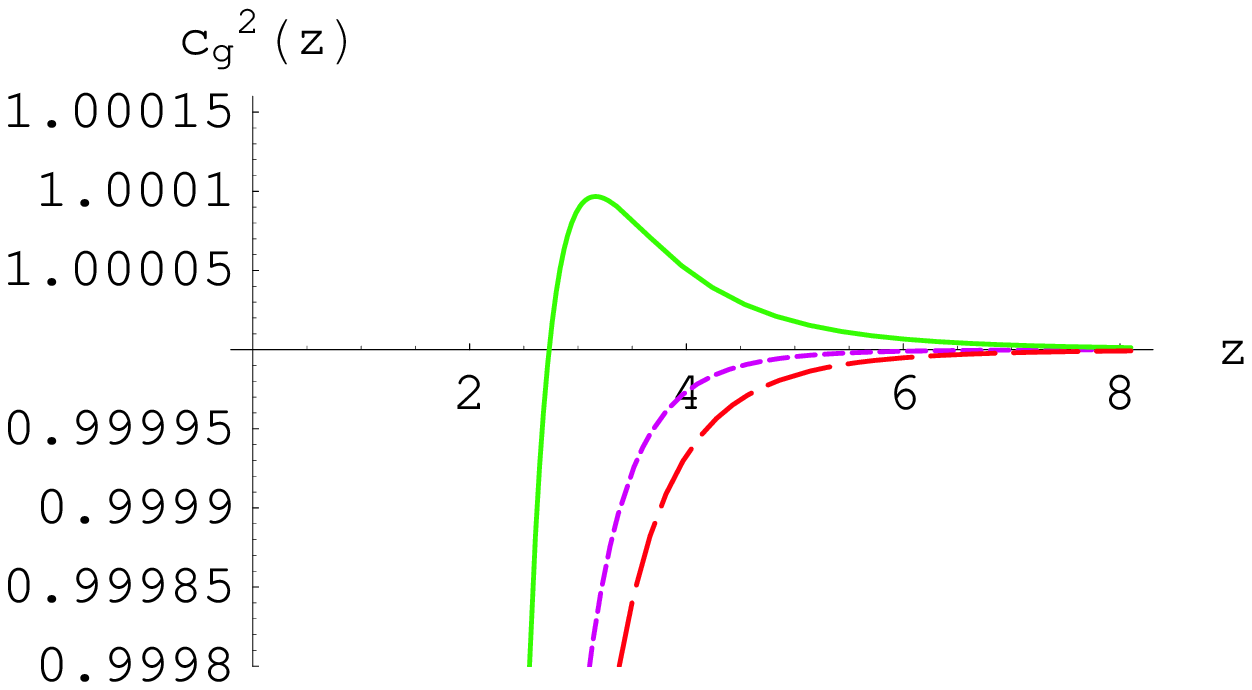,height=2.1in,width=2.9in}
\end{center}
\caption{\label{fig4} The speed of graviton with $\lambda\Z{\rm
GB}=0.084$ (left plot) and $\lambda\Z{\rm GB}=0.09$ (right plot),
with $\tilde{\epsilon} =0.3$ (solid, green), $\tilde{\epsilon}=0$
(dash, purple) and $\tilde{\epsilon}=-0.3$ (long-dash, red). Here
it is assumed that $\tilde{\epsilon}/q^2 \ll 1/2 $ (i.e. $q\to
\infty$).}
\end{figure}

\begin{figure}[ht]
\begin{center}
\hskip-0.3cm
\epsfig{figure=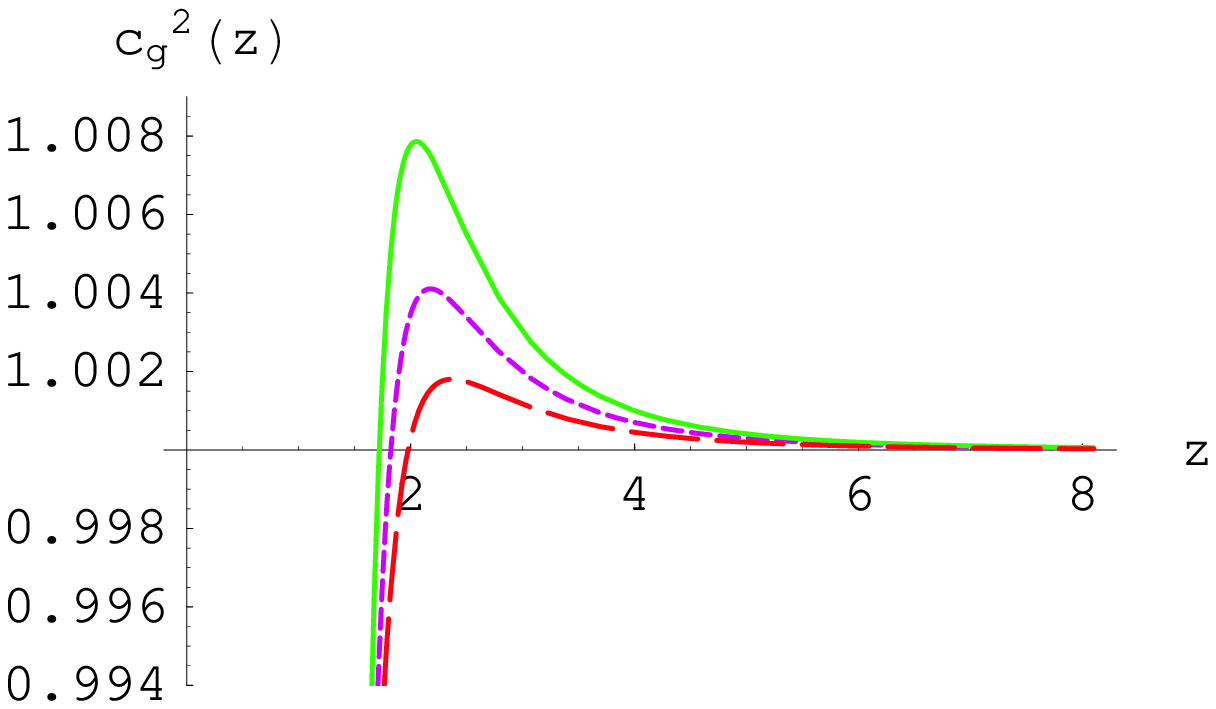,height=2.1in,width=2.9in}
\hskip0.2cm
\epsfig{figure=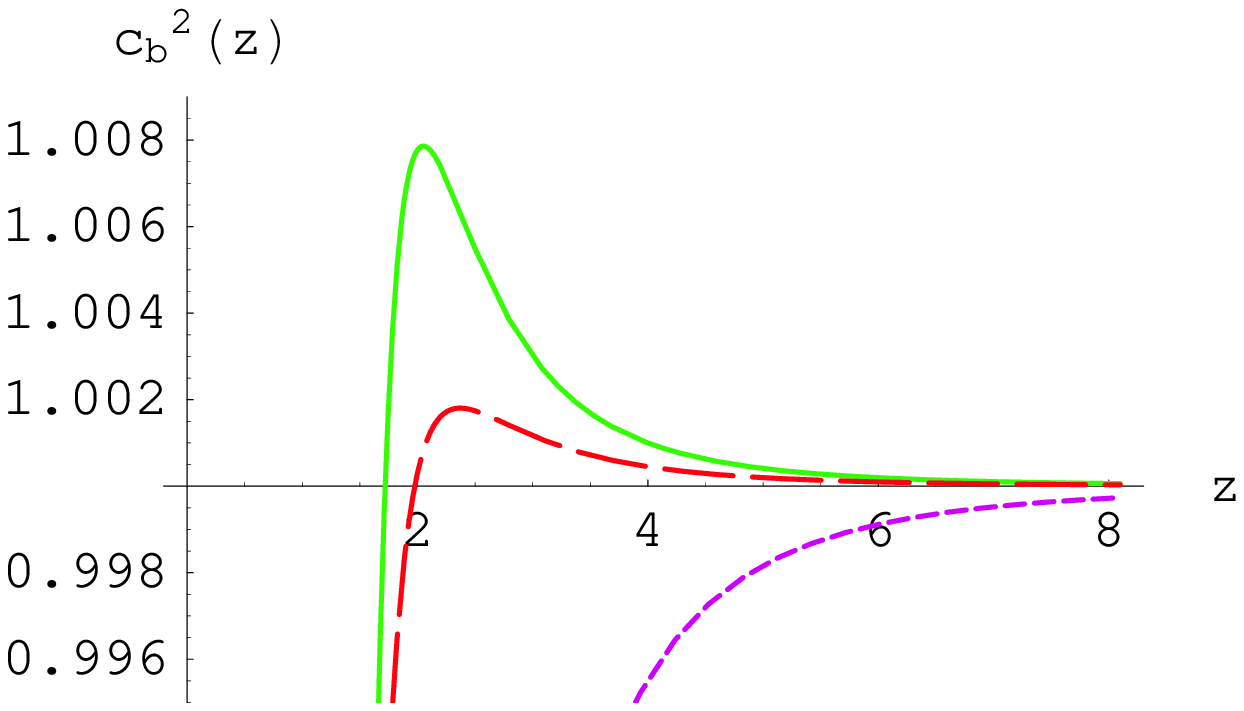,height=2.1in,width=2.9in}
\end{center}
\caption{\label{fig5} The speed of graviton $c_g^2$ (left plot)
and the local speed of light $b_g^2$ (right plot) with
$\lambda\Z{\rm GB}=0.1$, and $\tilde{\epsilon} =0.3$ (solid,
green), $\tilde{\epsilon}=0$ (dash, purple),
$\tilde{\epsilon}=-0.3$ (long-dash, red).}
\end{figure}

The effect of curvature at $z\to \infty$ can be known also by
studying the propagation speed of tensor modes on a constant
$r$-hypersurface. In particular
\begin{equation}
c_g^2(z)\equiv \frac{N_*^2 \tilde{f}}{z^2}\,\frac{(1-\lambda\Z{\rm
GB}
\partial_z^2\tilde{f})(1+2\tilde{\epsilon}/\tilde{q}^2)}{1-\lambda\Z{\rm
GB}\partial_z \tilde{f}/z}
\end{equation} can be interpreted as the square of local speed
of graviton on a constant $r$-hypersurface, by identifying
$N_*\equiv a$. The speed of graviton increases with the strength
of the coupling $\lambda\Z{\rm GB}$; this increase is maximum for
a dual field theory defined on $\epsilon=+1$ hypersurfaces (see
figure~\ref{fig4}). In the $\epsilon=0$ case, the square of local
speed of light, $c_b^2= \frac{N_*^2 \tilde{f}}{z^2}$, is always
sub-luminal, but it can be super-luminal for $\epsilon \ne 0$ (see
figure~\ref{fig5}). This is due to a well known fact that in the
presence of higher curvature terms the graviton wave packets in
general do not propagate on the light cone of a given background
geometry.

\subsection{Black Hole Thermodynamics and Entropy Bound}

We now briefly discuss the thermodynamics properties such as free
energy and entropy of AdS Gauss-Bonnet black holes. Indeed,
spherical AdS black holes present a different qualitative feature,
namely a phase transition at finite temperatures. This can be
seen, for instance, by considering the free energy of a
spherically symmetric AdS$_5$ GB black hole. In $d$-spacetime
dimensions, the free energy is given by (see, e.g.,~\cite{Ish03b})
\begin{equation}
{F}=\frac{V_{d-2} (x L)^{d-3}}{16\pi G_N} \left[(\epsilon-x^2)+
\frac{(d-2)}{(d-4)} \frac{\lambda_{\rm GB}\epsilon^2}{x^2}
-\frac{(d-1)}{(d-4)}\frac{\epsilon \lambda_{\rm GB}(2
x^2+\epsilon)} {x^2+ \epsilon\lambda_{\rm GB}}\right]-M_{\rm extr}
\delta_{\epsilon,-1},
\end{equation}
where the extremal ADM (Arnowitt-Deser-Misner) mass $M\Z{\rm
extr}$, which is nonzero for $\epsilon=-1$, is given by
\begin{equation}
M_{\rm extr}\equiv \frac{(d-2) V_{d-2} L^2}{16\pi G_N} \,\mu\Z{\rm
extr} = \frac{(d-2) V_{d-2}}{16\pi G_N}L^{d-3}
\left[(\epsilon+x^2)\, x^{d-3}+\epsilon^2 \lambda_{\rm GB}
x^{d-5}\right]|\Z{x=x\Z{\rm extr}},
\end{equation}
where, as above, $x\equiv r\Z{+}/L$. For a definiteness, we shall
take $d=5$. This choice is again motivated from AdS/CFT
correspondence~\cite{Maldacena97a}. In the limit $x\to 0$, we find
\begin{equation}
F\Z{\epsilon=+1} =\frac{3 V_3 L^2}{16\pi G_N} \lambda_{\rm GB}.
\end{equation}
This corresponds to the free energy of a boundary field theory
defined on $S^3$. However, for $\epsilon=0$ and $\epsilon=-1$, the
free energy always vanishes in the limit $x\to 0$. In $d=5$, and
with $\lambda\Z{\rm GB}<1/4$, the total thermodynamic
energy~\cite{Ish03b}
\begin{equation}
E= T {\cal S} + F=\frac{3 V_3 L^2}{16\pi G_N}\left[x^4+ \epsilon
x^2+ \lambda_{\rm GB} \epsilon^2+  \frac{1}{4}
\left(1-4\lambda\Z{GB}\right)\,\delta_{\epsilon,-1}\right]
\end{equation}
is a positive concave function of the black hole's temperature for
all values of $\epsilon$ ($=0, \pm 1$).

Next, we consider the change in entropy, which is given by
\begin{equation}
\frac{d {\cal S}}{dr_+}=\frac{(d-2)V\Z{d-2} r\Z{+}^{d-1}}{4\pi
G_N} \left(1+\frac{2\lambda\Z{\rm GB}{\epsilon}}{x^2}\right).
\end{equation}
With $\lambda\Z{\rm GB}>0$, $d{\cal S}/dr_+$ is always positive
for $\epsilon\ge 0$. In fact, $d{\cal S}/dr_+$ is positive also
for $\epsilon=-1$, since the minimal horizon of a GB black hole is
$x\Z{\rm min}^2 > -2\epsilon \lambda\Z{\rm GB}$. Nevertheless, at
small distances, like $x< \sqrt{1/2}$, the temperature of a
spherically symmetric Gauss-Bonnet black hole becomes larger than
that of a Schwarzschild black hole, at a given (fixed) entropy,
indicating a classical instability of the system. An earlier study
by Odintsov and Nojiri in~\cite{Nojiri:02q} showed that the
negative entropy state corresponds to the maximum rather than to
the minimum of Euclidean action or free energy. Although the
discussion in~\cite{Nojiri:02q} was restricted to $c=0$ in
eq.~(\ref{main-action}), similar arguments hold for $c\ne 0$, or a
nonzero $\lambda_{GB}$.

When $\epsilon=-1$, there exists an extremal solution with zero
Hawking temperature at $r\Z{+}=r\Z{\rm extr}=\frac{L}{\sqrt{2}}$
and $\mu\Z{\rm extr}=\frac{L^2}{4} (4\lambda\Z{\rm GB}-1)$. It is
easy to check that
\begin{equation}
{\cal S}|\Z{T\to 0} =\frac{V_3}{G_N}
\frac{L^3}{2^{7/2}}\left(1-12\lambda\Z{\rm GB}\right), \qquad
E|\Z{T\to 0}=0.
\end{equation} The positivity of
entropy requires $\lambda\Z{\rm GB}<1/12$; when $\lambda_{\rm GB}
> 1/12$, the solutions may violate the second law (of black
hole thermodynamics),
rendering the theory inconsistent. This observation is consistent
with a recent discussion in~\cite{Brigante-etal}, where, for a
class of $3+1$-dimensional CFTs in flat space, the GB gravity
violates a micro-causality for $\lambda_{\rm GB}
> 0.9$; particularly, the local speed of graviton can exceed the
speed of light in this limit.

We also note that, for a $(3+1)$-dimensional CFT duals of
$4+1$-dimensional Gauss-Bonnet gravity with $\epsilon=-1$,
consistency of the theory may require
\begin{equation}
\lambda\Z{\rm GB} < 1/12, \qquad \frac{\eta}{s} \ge 2/3
\left(\frac{1}{4\pi}\right),
\end{equation}
which puts 2.6\% weaker limit on $\eta/s$ than
in~\cite{Brigante-etal} but 33.4\% stronger limit than the KSS
bound. When $d=6$ and $d=7$, the consistency of Gauss-Bonnet
gravity requires, respectively,
\begin{equation}
\lambda\Z{\rm GB}^{d=6} \le 0.1380, \qquad \lambda\Z{\rm GB}^{d=7}
\le 0.1905.
\end{equation}
Similar results hold with $({\rm Riemann} )^2$ corrections to
classical Einstein gravity.

\section{Entropy Bound for (Riemann)$^2$ Gravity}

Another interesting theory of extended gravity is obtained by
setting $a=b=0$ in eq.~(\ref{main-action}). When $d=5$, the action
(\ref{main-action}) corresponds to an effective AdS$_5$ action
deduced from a 10d heterotic string theory via heterotic type I
duality~\cite{Narain99a}
\begin{equation}\label{Riemann-action}
I=\frac{N_c^2}{4\pi^2 {L}^3} \int d^5x \left(R+\frac{12}{L^2}+
\frac{{L}^2}{16N_c}\,R_{\mu\nu\lambda\rho}
R^{\mu\nu\lambda\rho}\right),
\end{equation}
where, using the AdS/CFT dictionary, the coefficient of $({\rm
Riemann})^2$ term has been fixed as $\alpha^\prime L^2=L^2/(16
N_c)\equiv \lambda\Z{\rm Riem} L^2/2$ and $N_c^2\equiv \pi
L^3/4G_N$ for AdS$_5$. Although, at a linearised level, the
effective theory defined by~(\ref{Riemann-action}) behaves
differently from that defined by Gauss-Bonnet action, for
instance, in terms of effective degrees of freedom or graviton
propagators, they both modify the corresponding dual field theory
variables including the ratio $\eta/s$ in a qualitatively similar
way.

To be specific, we consider the black hole entropy and Hawking
temperature of a $({\rm Riemann})^2$ corrected AdS$_5$ black hole,
which are given by~\cite{Ish02} (see
also~\cite{Nojiri:01a,Dutta:2006})
\begin{equation}\label{Riem-entropy}
{\cal S} = \frac{A}{4G_N} \left[1+2\lambda\Z{\rm Riem}
\left(2+3\tilde{\epsilon}\right)+{\cal O}(\lambda\Z{\rm
Riem}^2)\right],
\end{equation}
\begin{equation}
 T =\frac{x}{2\pi L}\left[
2+\tilde{\epsilon} -2\lambda\Z{\rm Riem} x^2
(1+\tilde{\epsilon}\,)^2+{\cal O}(\lambda\Z{\rm Riem}^2)\right],
\end{equation}
where $A\equiv V_3 \,r\Z{+}^3$ and, as above,
$\tilde{\epsilon}\equiv \epsilon/x^2$. Note that, with
$|\lambda\Z{\rm Riem}|> 0$, the entropy is not simply given by
one-quarter the area even for black brane solutions
($\epsilon=0$), indicating a breaking of conformal symmetry of the
boundary field theory. This also implies that a bulk gravity other
than with Einstein-Hilbert and Gauss-Bonnet gravity actions may
not have a well defined boundary CFT dual; with $({\rm
Riemann})^2$ terms, the boundary conditions alter the symmetry
algebras of a pure AdS space or of the general relativity.

Following the perturbative metric solution found in~\cite{Ish02}
and the methods discussed in
refs.~\cite{Brigante-etal,Kats:2007mq} for calculating shear
viscosity, we find
\begin{equation}
\eta=\frac{r\Z{+}^3 N_c^2}{4\pi^2 L^3} \left(1-4\lambda\Z{\rm
Riem}+{\cal O}(\lambda\Z{\rm Riem}^2)\right).
\end{equation}
The entropy density of a $({\rm Riemann})^2$ corrected black hole
is given by~\cite{Ish02,Ish04C}
\begin{equation}
s=\frac{r_+^3 N_c^2}{\pi L^3} \left(1+4\lambda\Z{\rm
Riem}
\left(1+\frac{3}{2}\tilde{\epsilon}\right)+{\cal
O}(\lambda\Z{\rm Riem}^2)\right).
\end{equation}
From these results we can infer that for a class of boundary field
theories in flat space ($\epsilon=0$) with $({\rm
Riemann})^2$-gravity dual, the ratio $\eta/s$ (up to the linear
term in $\lambda\Z{\rm Riem}$) is
\begin{equation}
\frac{\eta}{s}=\frac{1}{4\pi}\left( \frac{1-4\lambda\Z{\rm
Riem}}{1+4\lambda\Z{\rm Riem}}\right)\approx
\frac{1}{4\pi}\left(1-8\lambda\Z{\rm Riem}\right).
\end{equation}
Our estimation of $\eta/s$ is different from that
in~\cite{Brigante-etal} by a factor of $2$~\footnote{The source of
this difference is that in our discussion the entropy of a $({\rm
Riemann})^2$-corrected black hole is not given by one-quarter the
area even in the $\tilde{\epsilon}=0$ case; the coupling
$\lambda\Z{\rm Riem}$ here is related to the coefficient of $({\rm
Riemann})^2$ term ($\alpha\Z{3}$) introduced
in~\cite{Brigante-etal} via $\lambda\Z{\rm Riem}=2\alpha\Z{3}$.}
but it agrees with the final result (eq. (5.6))
in~\cite{Kats:2007mq}.

\begin{figure}[ht]
\begin{center}
\hskip-0.3cm
\epsfig{figure=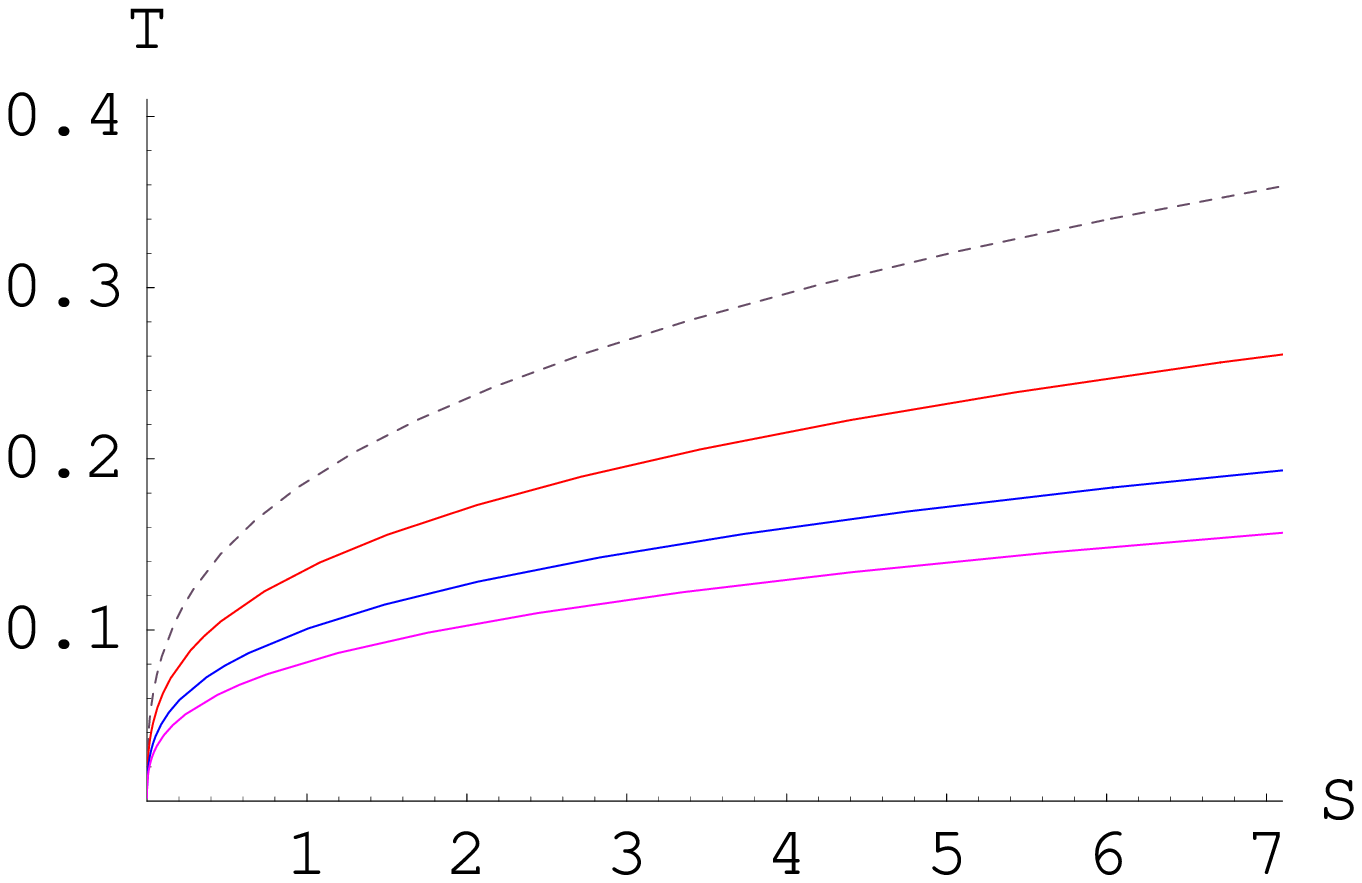,height=2.1in,width=2.9in}
\hskip0.2cm
\epsfig{figure=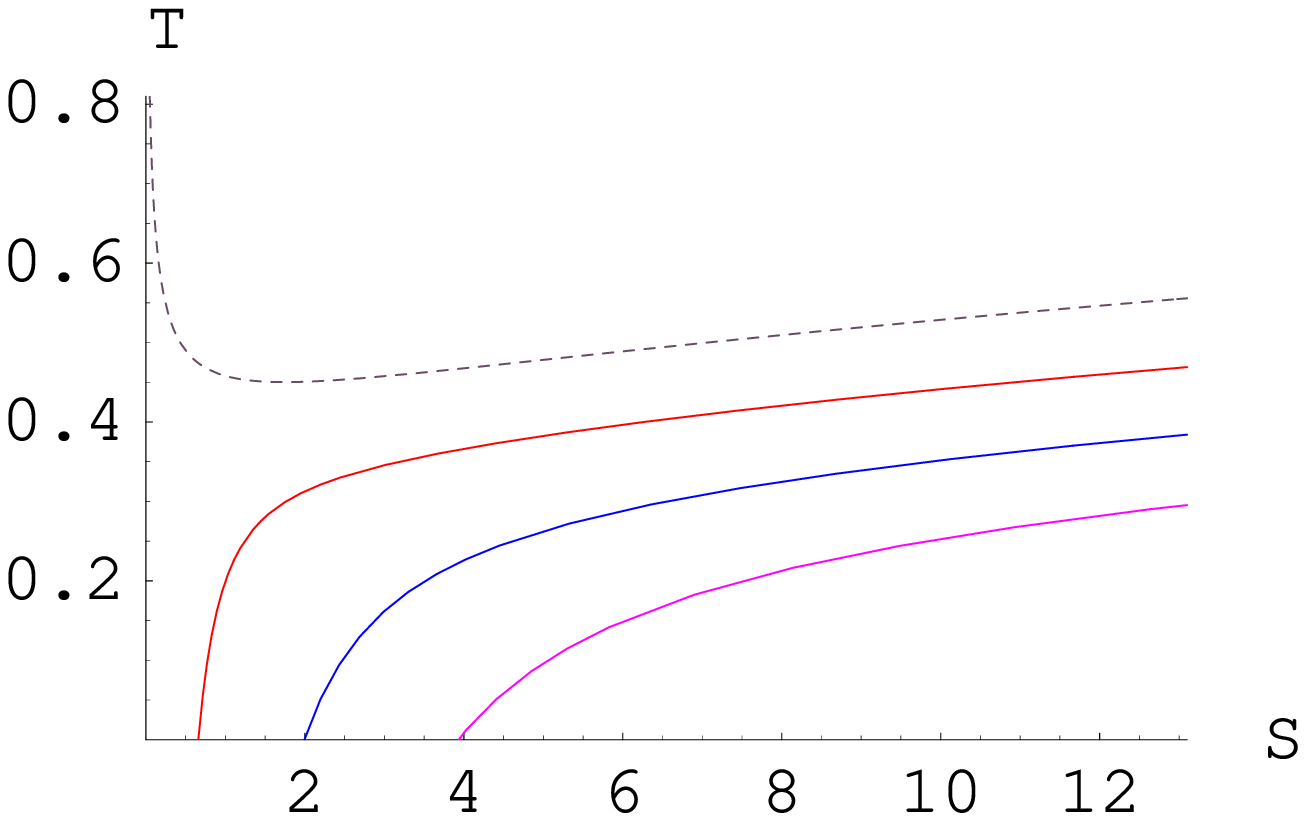,height=2.1in,width=2.9in}
\end{center}
\caption{\label{fig6} Hawking temperature vs entropy. Left plot:
$\epsilon=0$ and $\lambda\Z{\rm Riem}=0, 0.15, 0.3, 0.4$ (top to
bottom). Right plot: $\epsilon=+1$ and $\lambda\Z{\rm Riem}=0,
0.05, 0.1, 0.15$ (top to bottom).}
\end{figure}

\begin{figure}[ht]
\begin{center}
\hskip-0.3cm
\epsfig{figure=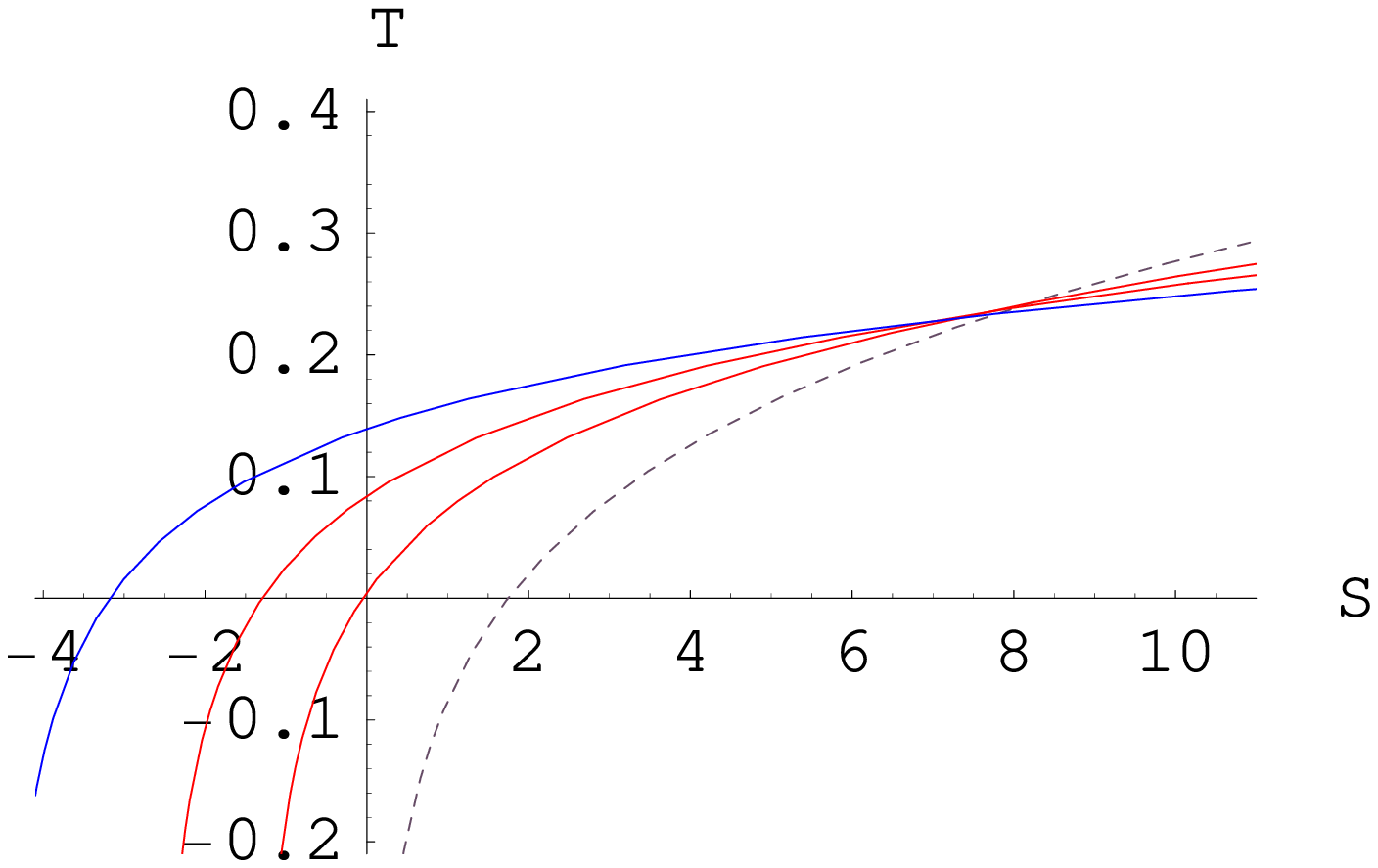,height=2.1in,width=2.9in}
\hskip0.2cm
\epsfig{figure=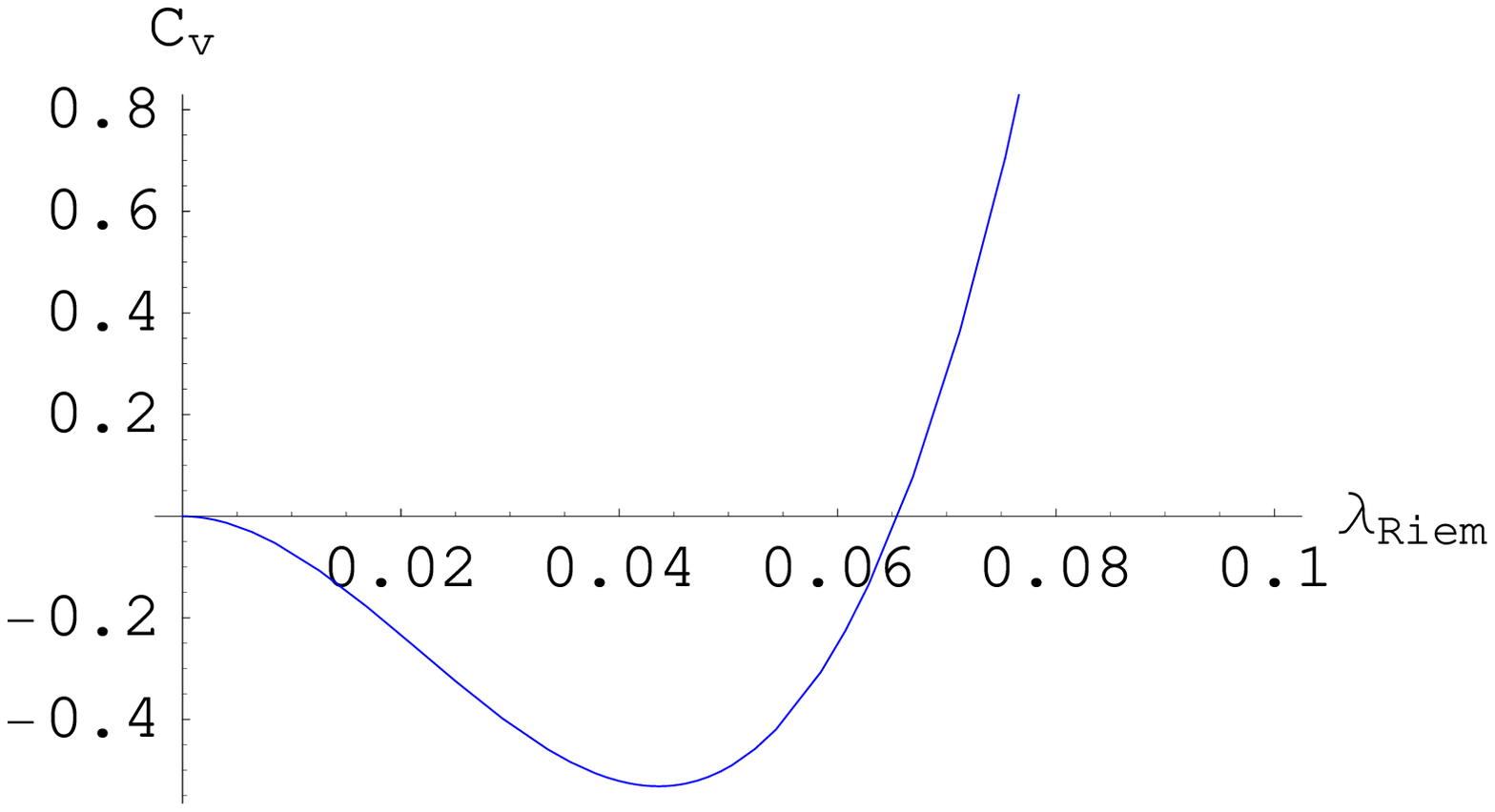,height=2.1in,width=2.9in}
\end{center}
\caption{\label{fig7} (Left plot) Entropy vs Hawking temperature
with $\epsilon=-1$ and $\lambda_{\rm Riem}=0.4, 0.25, 0.15, 0$
(left to right). (Right plot) Specific heat (at the extremal
state) as a function of $\lambda\Z{\rm Riem}$.}
\end{figure}

In the $\epsilon=0$ and $\epsilon=+1$ cases, the Hawking
temperature of a $({\rm Riemann})^2$-corrected black hole is
always found to be less than that of a Schwarzschild black hole,
at a given entropy. Also, the black hole entropy is non-negative
in the limit $T\to 0$ (see figure~\ref{fig6}). In the
$\epsilon=-1$ case, however, the entropy becomes negative at zero
Hawking temperature, especially, for $\lambda\Z{\rm Riem}
> 0.15$ (see also figure~\ref{fig7}), indicating a possible
violation of the second law of thermodynamics.

With $\epsilon=-1$, the Hawking temperature vanishes at $x\simeq
\sqrt{\frac{1}{2}(1+ \lambda\Z{\rm Riem})}$. The positivity of
extremal black hole entropy density
\begin{equation}\label{Riemann-extremal-s}
{s}\Z{(\rm AdS)}|\Z{T\to 0}=\frac{1}{G_N}
\frac{L^3}{2^{7/2}}\left(1-8\lambda\Z{\rm Riem}\right),
\end{equation}
which now requires $\lambda\Z{\rm Riem} <1/8$, ensures that the
total thermodynamic energy~\cite{Ish04C}
\begin{equation}
E=\frac{3 V_3 L^2}{16\pi G_N}\left[x^2(\epsilon+x^2)+\lambda\Z{\rm
Riem} (\epsilon+x^2)(\epsilon+3 x^2)+\frac{1}{4}(1+\lambda\Z{\rm
Riem})\right]
\end{equation}  is non-negative.
For a class of CFTs in flat space with $({\rm Riemann})^2$ gravity
dual, we find
\begin{equation}
\frac{\eta}{s}=\frac{1}{4\pi} \left(\frac{N_c-1/2}{N_c+1/2}\right)
\simeq \frac{1}{4\pi}\left(1-\frac{1}{N_c}\right).
\end{equation}
In the limit $N_c\to \infty$, one has $\eta/s \approx 1/4\pi$, but
$\eta/s<1/4\pi$ for a finite $N_c$. However, it does not mean that
for a class of CFTs with $({\rm Riemann})^2$ gravity dual, the
ratio $\eta/s$ is always smaller than $1/4\pi$. In fact, the
minimum of entropy density (and hence the maximum of $\eta/s$)
occurs for the $\epsilon=-1$ extremal solution at $x\simeq
\sqrt{(1+\lambda\Z{\rm Riem})/2}$~\footnote{Unlike for
Gauss-Bonnet gravity, the location of the extremal horizon of a
$({\rm Riemann})^2$ corrected AdS$_5$ black hole depends on the
coupling $\lambda\Z{\rm GB}$.}, which is given by
\begin{equation}
\frac{\eta}{s}= \frac{1}{4\pi} \left(1+4\lambda\Z{\rm Riem}-{\cal
O}(\lambda\Z{\rm Riem}^2)\right).
\end{equation}
Taking into account all three possibilities for the boundary
topology that $\epsilon=0$ or $\epsilon=\pm 1$, we find that
consistency of $({\rm Riemann})^2$ gravity requires
\begin{equation}
0 < \frac{\eta}{s}\le \frac{3}{2} \left(\frac{1}{4\pi}\right).
\end{equation}
It is an open question whether either of these limits applies to
nuclear matter at extreme densities and temperatures, or heavy ion
collision experiments exhibiting perfect fluid behaviour.
Nevertheless, it is intriguing that a general consideration based
on AdS black hole solutions with curvature square corrections
provide such bounds on $\eta/s$.

We end this section with a couple of remarks. In particular, in
the AdS$_5 \times S^5$ string background dual to ${\cal N}=4$
$SU(N_c)$ SYM gauge theory at strong coupling ($g\Z{\rm YM}^2 N_c
\to \infty$), the ratio $\eta/s$ has been found to increase once
the ${\alpha^\prime}^3 R^4$ terms are added in the effective
action of type IIB string theory, as implied
by~\cite{Buchel:2004,Myers-08b}
\begin{equation}
\frac{\eta}{s}=\frac{1}{4\pi} \left(1+ \frac{15\,
\zeta(3)}{\lambda^{3/2}}+\cdots \right),
\end{equation}
where $\zeta(3)=1.0202$. Even if this result is correct (including
the precise coefficient and the sign of the correction), the
conjectured KSS bound $\eta/s\ge 1/4\pi$ may be violated at a
finite coupling, since the ${\alpha^\prime}^3 R^4$ type
corrections are suppressed relative to the $1/N_c$ corrections
arising from $({\rm Riemann})^2$ terms. One may argue that for a
particular version of string theory, for example, type IIB string
theory, the $R^2$ type corrections could be absent due to some
supersymmetric conditions and hence the KSS bound $\eta/s\ge
1/4\pi$ may hold. Nevertheless, one cannot deny the role of $R^2$
type corrections in a full quantum theory of gravity, with all
possible higher derivative and higher-order curvature
contributions. It would be interesting to check a universality of
the result like $\eta/s\approx 1/4\pi$ through numerical
hydrodynamic simulations of data from RHIC and LHC.

\section{Conclusion}

The Gauss-Bonnet gravity with a coupling $\lambda\Z{GB}<1/4$ could
be viewed as a classical limit of a consistent theory of quantum
gravity. Such a theory is consistent with the prediction of some
low energy effective superstring models or the supergravity
approximation of string theory. In a full quantum theory of
gravity, with all possible higher derivative and higher-order
curvature contributions, though, we cannot find such explicit
bounds for the couplings
--- the solutions can be known only at perturbative level.
We must instead find new conditions strong enough to ensure the
presence of a black hole, but weak enough to be allowed by
positivity of (extremal) entropy and the total thermodynamic
energy.

It is conceivable that the bound on the shear viscosity of any
fluid in terms of its entropy density is saturated,
$\eta/s=1/4\pi$, for gauge theories at large 't Hooft coupling,
which correspond to the cases where all higher-order curvature
contributions are absent. But this bound is naturally in immediate
threat of being violated in the presence of generic higher
derivative or higher-order curvature corrections to the
Einstein-Hilbert action. It is remarkable that by tuning of the
Gauss-Bonnet coupling, the ratio $\eta/s$ can be adjusted to a
small positive value or even to zero.

In our work, limits on $\lambda\Z{GB}$ are imposed by demanding
the positivity of extremal black hole entropy and non-violation of
boundary causality. The results like~(\ref{GB-extremal-s}) may
receive nontrivial corrections in the presence of further higher
derivative terms, like quartic terms in Weyl tensors. Similar
remarks could be made in regard to~(\ref{Riemann-extremal-s}).
However, this kind of reasoning cannot be used to minimize or dim
the significance of extended gravity theories with Gauss-Bonnet or
$({\rm Riemann})^2$ terms alone. We have not found any obvious
explicit bound on $\lambda\Z{GB}$ from the thermodynamics of
spherically symmetric AdS Gauss-Bonnet black holes, which may
however arise as a consequence of boundary causality. What we find
really interesting is that a critical value of $\lambda\Z{GB}$
beyond which the theory becomes inconsistent is related to the
entropy bound for a class of AdS GB black holes with a hyperbolic
or Euclidean anti-de Sitter event horizon. We have come to similar
outcomes for Riemann squared gravity. In our discussions,
inconsistency of the (Gauss-Bonnet and Riemann squared gravity)
theory, such as a violation of micro-causality, has been shown to
be related to a classical limit on black hole entropy.

Our results also indicate an interesting connection between the
thermodynamics of black hole horizons and the quadratic Euler
invariants present in extended theories of gravity, where the
relations ${\cal S}=|\beta|(E-F)$ and $d{\cal S}=\beta dE$ may
have a greater domain of validity than that of classical Einstein
gravity, see also ref.~\cite{Paddy:2007a} for related discussions.
An interesting open question is whether Hawking radiation and
black hole evaporation can also fit into extended gravity
theories. It would be interesting to know how generic higher
derivative corrections modify various (dual) gauge theory
observables both at finite and strong coupling limits. If similar
bounds on coupling constants can be found for spinning AdS-GB
black holes and/or generalised Gauss-Bonnet black holes in the
presence of Maxwell charges or electromagnetic field strengths in
the matter action, it would represent an important progress.

In~\cite{Ge-Sin:08b}, it has just been reported that, in the
presence of a GB term and a Maxwell type charge $q$, the ratio
$\eta/s$ is given by $4\pi (\eta/s)= 1-4\lambda\Z{\rm GB}(1-a/2)$,
where $a\equiv q^2 L^6/r\Z{+}^2$. In the extremal limit ($a\to
2$), one can restrict $\lambda\Z{\rm GB}$ such that $\lambda\Z{\rm
GB}\le 1/24$, the latter ensures that the gravitational potential
of a black brane is positive and bounded. The bound $4\pi (\eta/s)
\ge 5/6$ reported in~\cite{Ge-Sin:08b} for a nonzero charge is
somewhat stronger than for pure EGB gravity in flat space, namely
$4\pi (\eta/s) \ge 2/3$.

One of the main motivations for studying higher-curvature or
higher-derivative corrected AdS black holes is gauge
theory-gravity duality or more generally the implications of AdS
black holes to dual CFTs as well as the thermal transport
properties of low energy QCD; the latter provide a real-time
dynamics at strong coupling limits. It is perhaps true that limits
on higher curvature couplings can be placed also by studying de
Sitter (black hole) spacetimes, but it is not understood yet
whether any de Sitter GB black holes would find a direct
application to real time dynamics of dual QFTs. Studying the GB
black holes in de Sitter (dS) spaces has some significance in its
own right. Indeed, for de Sitter GB blacks holes, there could
arise a cosmological event horizon in addition to a black hole
horizon. In turn, one would expect a separate set of thermodynamic
variables for the cosmological horizon. In general, the set of
thermodynamics quantities, for instance, the entropy densities,
associated with the black hole horizon and cosmological horizon
are not equal. As a consequence, the spacetime for a GB black hole
in dS space may not be stable semi-classically. Also, an ambiguity
could arise as to whether the thermal transport properties of low
energy QCD correspond to that defined on cosmological or black
hole horizons. Nevertheless, we find some earlier discussions
in~\cite{Cvetic:01bk} particularly encouraging, which indicate
that the entropy of a de Sitter GB black hole become negative for
$\frac{1}{12} < \lambda\Z{\rm GB} < 1/4$, leading to a possible
violation of unitarity in this range. This suggests some deeper
connections between dS and AdS Gauss-Bonnet black hole solutions.

\vspace{1.5ex}
\begin{flushleft}
\large\bf Acknowledgments
\end{flushleft}

IPN acknowledges the hospitality of IUCAA where a part of the work
was carried out. The work of IPN is supported by the New Zealand
Foundation for Research, Science and Technology Grant No. E5229
and also by Elizabeth Ellen Dalton Grant No. 5393.

\appendix
\section{Appendix: Charged AdS Gauss Bonnet Black Holes}
\renewcommand{\theequation}{A.\arabic{equation}}
\setcounter{equation}{0}

Consider the $d$-dimensional Einstein-Gauss-Bonnet gravitational
action
\begin{equation}
I\Z{\rm g}=\frac{1}{16\pi G} \int d^d x\sqrt{-g}
\left[R+\frac{(d-1)(d-2)}{L^2}+ \alpha^\prime L^2\, {\cal R}^2
\right],
\end{equation}
where ${\cal R}^2=R^2-4R{ab} R^{ab}+R_{abcd} R^{abcd}$, along with
the standard Maxwell type action
\begin{equation}
I\Z{\rm EM}=-\frac{1}{4} \int d^d x \sqrt{-g} \,F_{a b} F^{ab}.
\end{equation}
In any $d$-dimensional AdS spacetimes admitting
$(d-2)$-dimensional Euclidean subspace
\begin{equation}
h^E = h_{\mu\nu}^E dx^\mu dx^\nu \equiv \frac{d\chi^2}{1-\epsilon
\chi^2} +\chi^2 d\Omega_{d-3}^2
\end{equation}
with the constant curvature $\epsilon=0, +1, -1$, the solution of
Einstein field equations following from the total action, $I\Z{\rm
g}+I\Z{\rm EM}$, is isometric to
\begin{equation}
ds^2 = -f(r) dt^2+ \frac{dr^2}{f(r)}+ r^2 h_{\mu\nu}^E dx^\mu
dx^\nu,
\end{equation}
with electric field strength
\begin{equation}
F\equiv \frac{q^2}{4\pi} r^{4-2d}\, dt\wedge dr
\end{equation}
and the metric potential
\begin{equation}
f(r)=\epsilon+\frac{r^2}{L^2}\frac{1}{2\lambda\Z{\rm GB}}
\left[1+\kappa \sqrt{1-4\lambda\Z{\rm GB}+\frac{4\lambda\Z{\rm
GB}L^2}{r^{d-1}}\left(\mu-Q^2 \,r^{3-d}\right)}\right],
\end{equation}
where $\lambda\Z{\rm GB}\equiv (d-3)(d-4)\alpha^\prime$ and
$\kappa=\pm 1$. We refer
to~\cite{Wiltshire:88u,Cvetic:01bk,Charmousis:08kc,Anninos:08sj}
for further discussions on black hole thermodynamic of charged
AdS-GB black holes; \cite{Anninos:08sj} presents a comprehensive
analysis for all three possible horizon topologies ($\epsilon=0$
or $\pm 1$).

The integration constants $\mu$ and $Q^2$ are related,
respectively, to the Arnowitt-Deser-Meisener (ADM) mass $M$ and
the Maxwell charge $q$ via
\begin{equation}\label{mass-formula}
M\equiv \frac{(d-2) V_{d-2}L^2}{16\pi G}\, \mu, \quad Q^2\equiv
\frac{q^2}{2\pi (d-2)(d-3)}.
\end{equation}
Needless to say, the Maxwell charge $q$ is topological in $d=3$.
The negative root solution ($\kappa=-1$) has a smooth limit as
$\lambda\Z{\rm GB}\to 0$, which is often referred to as the
`Einstein branch'. While the positive root solution ($\kappa=+1$),
which has no smooth limit as $\lambda\Z{\rm GB}\to 0$, represents
a distinct new feature of Gauss-Bonnet gravity which is completely
absent in pure Einstein gravity in any dimensions. As anticipated,
for the negative root solution ($\kappa=-1$), the metric potential
$\tilde{f}(z)\equiv f(r) L^2/r\Z{+}^2$ is negative inside the
black hole horizon ($z\equiv r/r_+ <1$) where the role of $t$ and
$r$ coordinates will interchange. Under a modulo double wick
rotation, we get
\begin{equation}
f(r)=\epsilon +\frac{r^2}{L^2}\frac{1}{2\lambda\Z{\rm GB}}
\left[1+\kappa \sqrt{1-4\lambda\Z{\rm GB}+\frac{4\lambda\Z{\rm
GB}}{r^{d-1}}\left(\mu+P^2\,r^{3-d}\right)}\right]
\end{equation}
with the magnetic field strength
\begin{equation}
F= -\frac{q^2}{4\pi} r^{4-2d}\, d\theta \wedge dr \equiv
\frac{g^2}{4\pi} r^{4-2d}\, d\theta \wedge dr
\end{equation}
where $g\equiv -iq$ is the magnetic charge and $P^2=-Q^2$.

\begin{figure}
\begin{center}
\hskip-0.3cm
\epsfig{figure=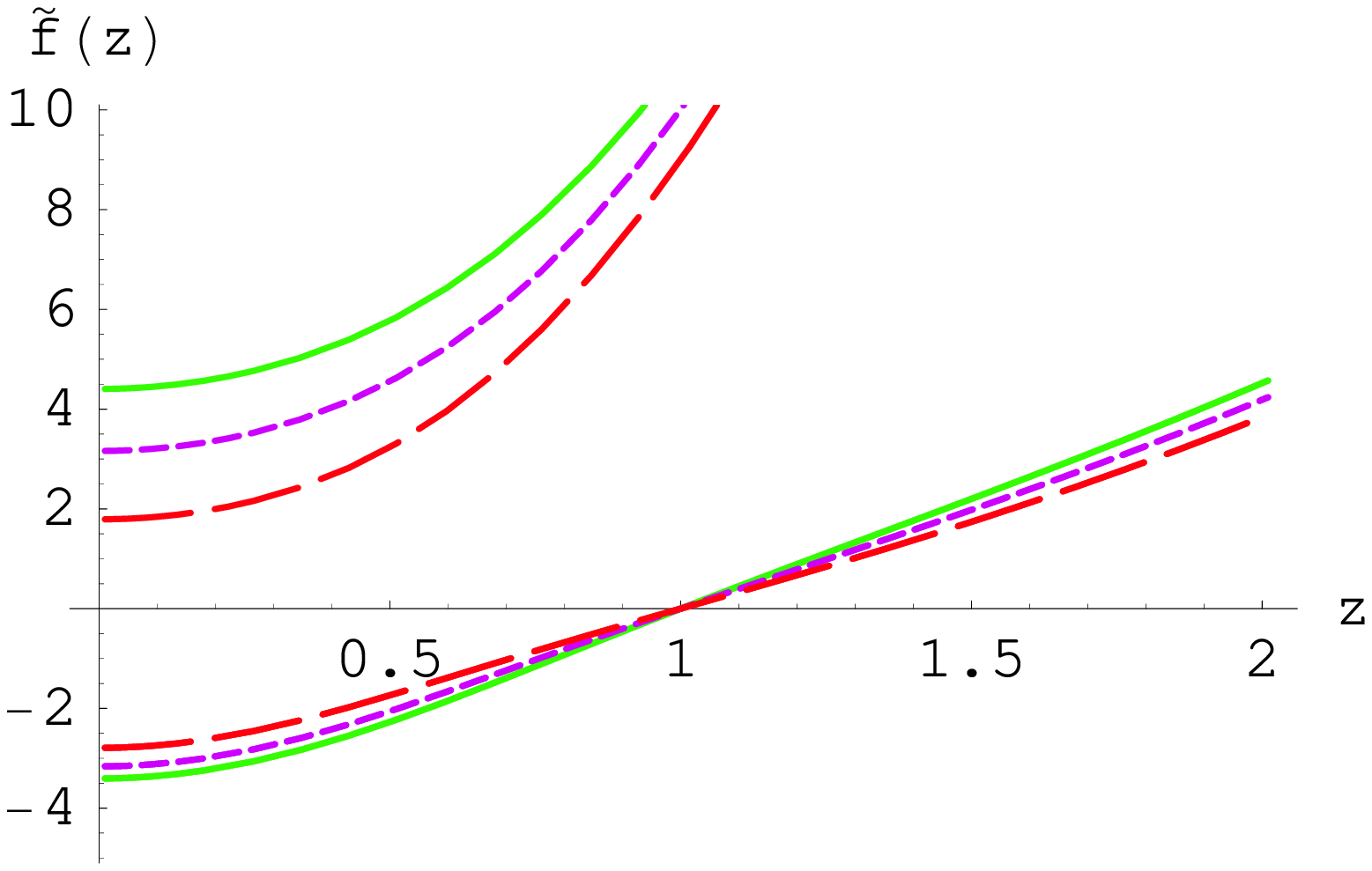,height=2.0in,width=2.9in}
\hskip0.2cm
\epsfig{figure=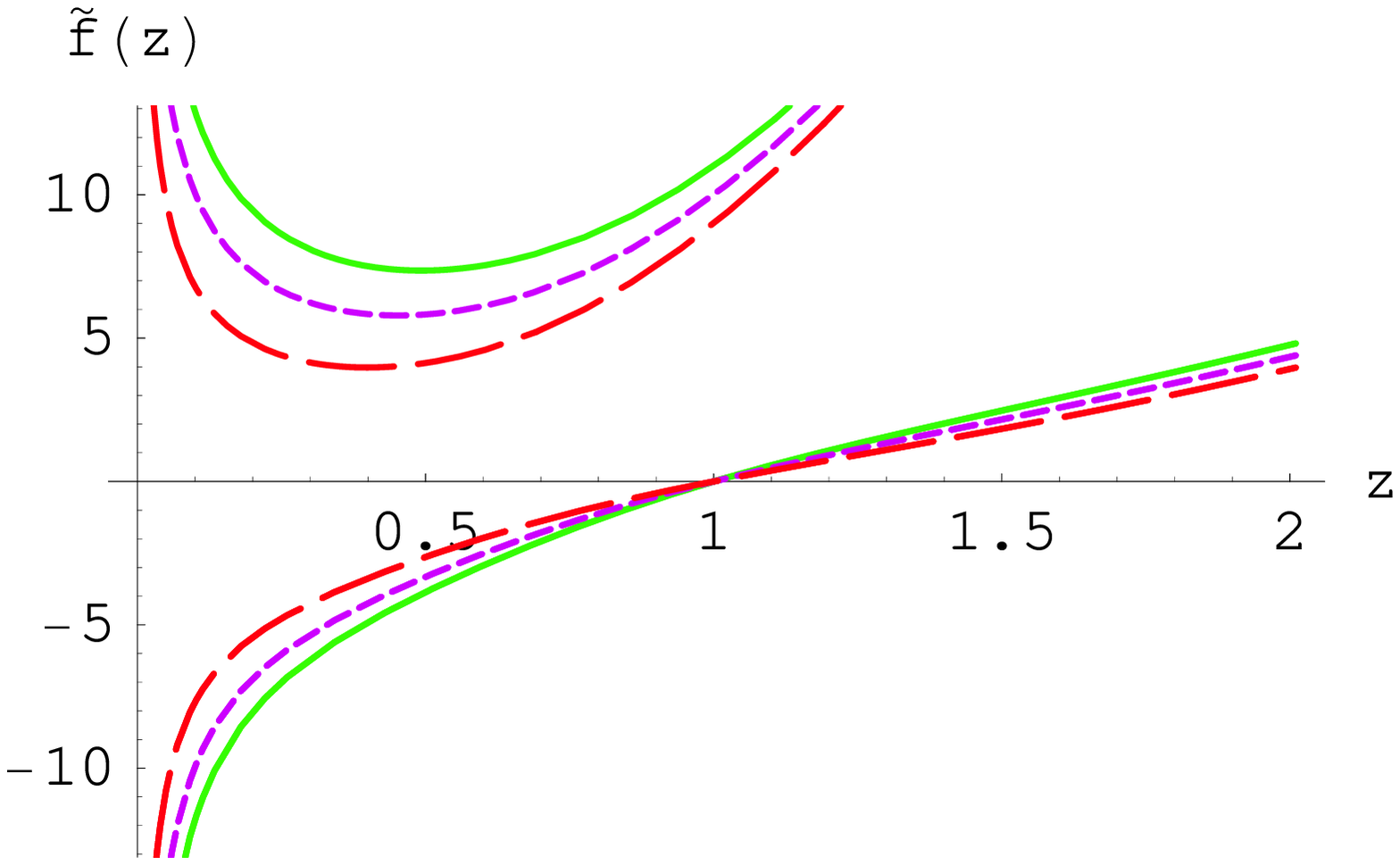,height=2.0in,width=2.9in}
\end{center}
\caption{\label{appen-fig1} The metric potential $\tilde{f}$ as a
function of $z$ with $Q^2=0$, $\lambda\Z{\rm GB}=0.1$, $d=5$ (left
plot), $d=6$ (right plot). The solid (green), short-dash (purple)
and long-dash (red) lines correspond, respectively, to
$\epsilon=+1$, $\epsilon=0$ and $\epsilon=-1$. We have taken
$x\equiv r_+/L=1$; making the horizon larger or smaller than the
AdS only shifts the height of $\tilde{f}(z)$. The top (bottom)
three lines correspond to $\kappa=+1$ ($-1$). In $d=5$,
$\tilde{f}(z)$ is regular for all values of $z$.}\end{figure}

\begin{figure}
\begin{center}
\hskip-0.3cm
\epsfig{figure=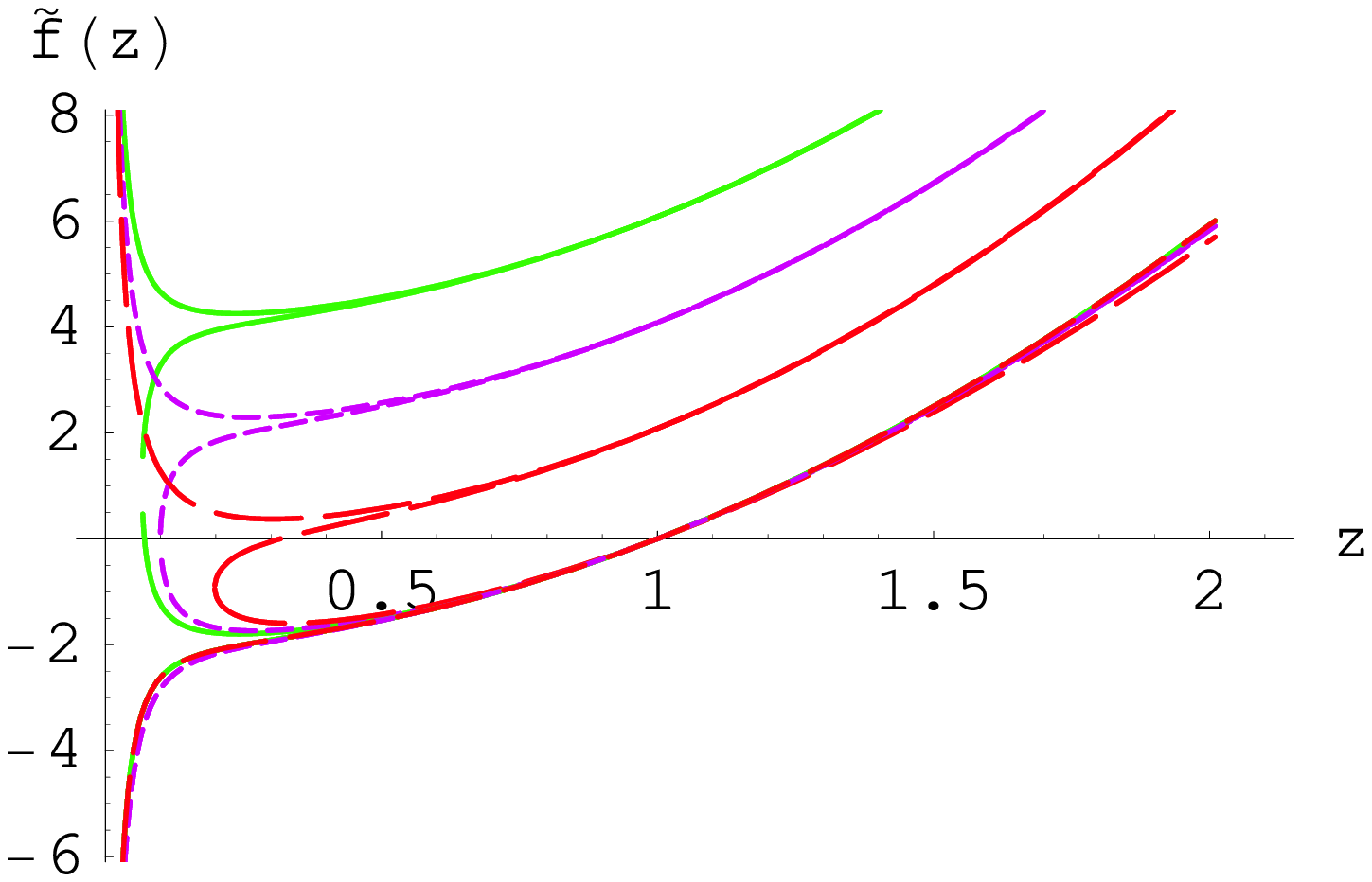,height=2.0in,width=2.9in}
\hskip0.2cm
\epsfig{figure=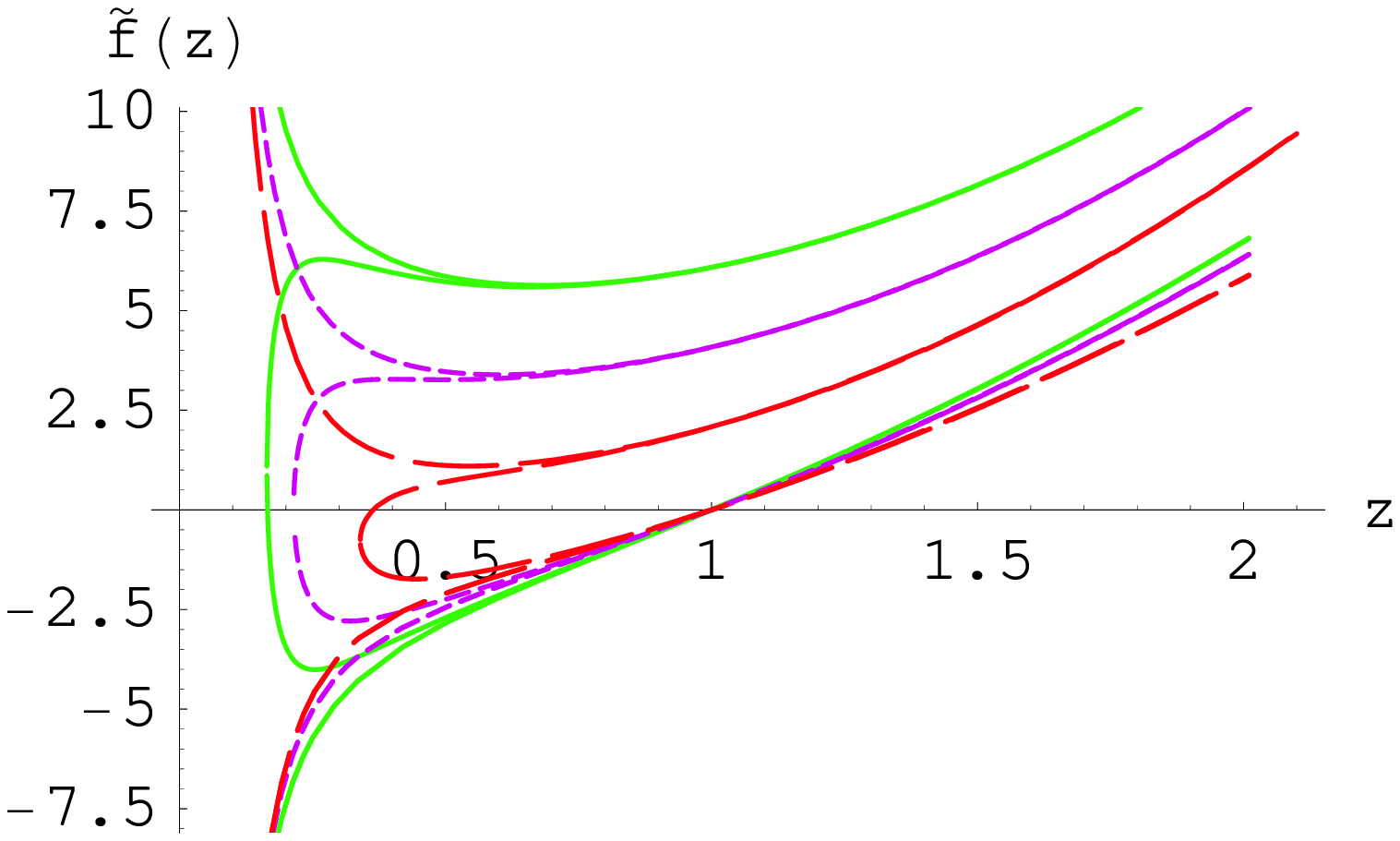,height=2.0in,width=2.9in}
\end{center}
\caption{\label{appen-fig2} The function $\tilde{f}(z)$ with
$\lambda\Z{\rm GB}=0.245$, $Q^2=0.01$ (or $P^2=0.01$), $d=5$ (left
plot) and $d=6$ (right plot). The solid (green), short-dash
(purple) and long-dash (red) lines correspond, respectively, to
$\tilde{\epsilon}=+1$, $\tilde{\epsilon}=0$ and
$\tilde{\epsilon}=-1$. For an electrically charged black hole, the
$\kappa=\pm 1$ branches join at $z\ll 1$, while, for a
magnetically charged black hole, they are disjoint; in the latter
case, $\tilde{f}(z)$ diverges to $+\infty$ ($-\infty$) for
$\kappa=+1$ ($\kappa=-1$) branch, as $z\to 0$.  }
\end{figure}

\begin{figure}
\begin{center}
\hskip-0.3cm
\epsfig{figure=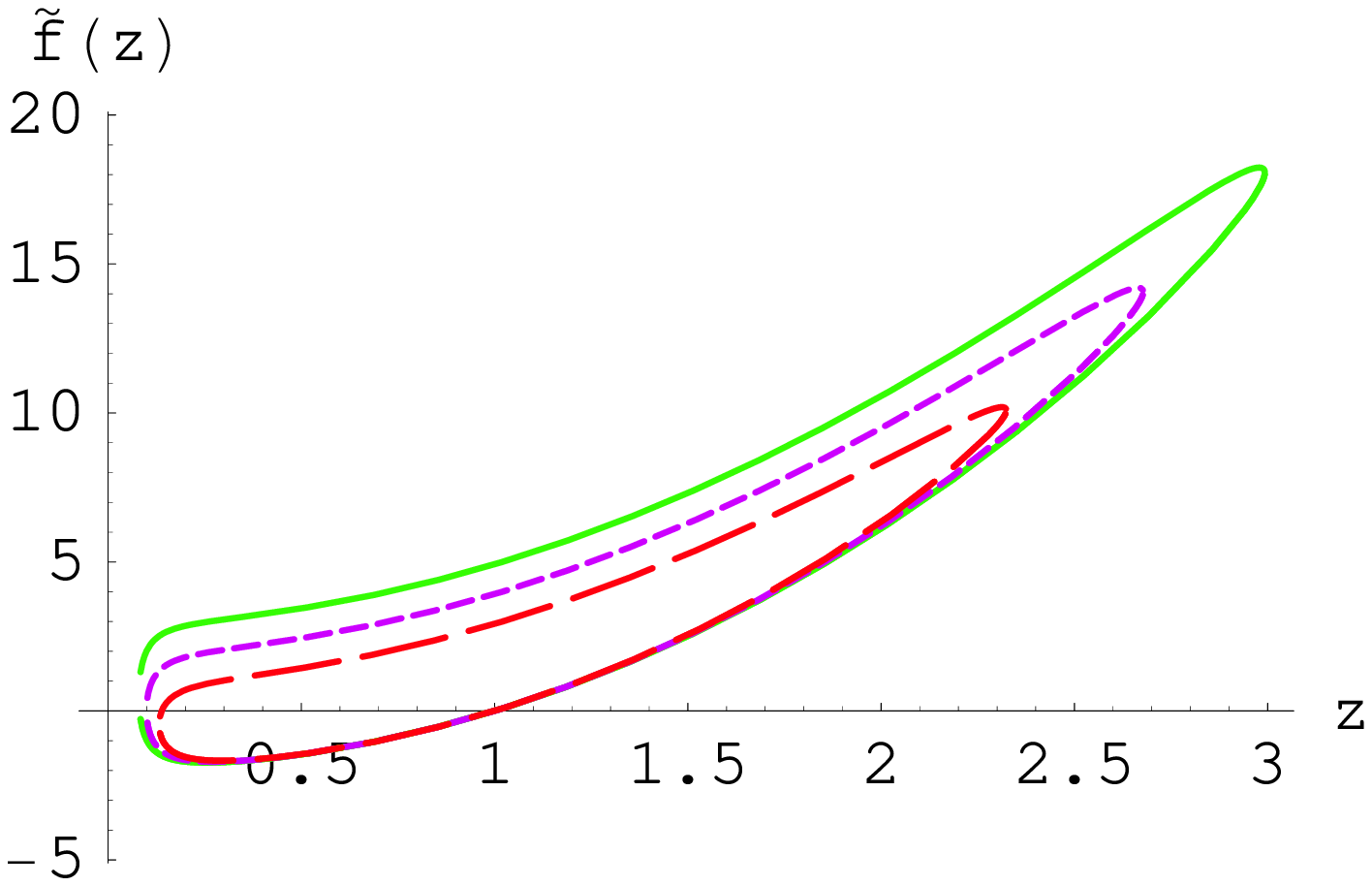,height=2.0in,width=2.9in}
\hskip0.2cm
\epsfig{figure=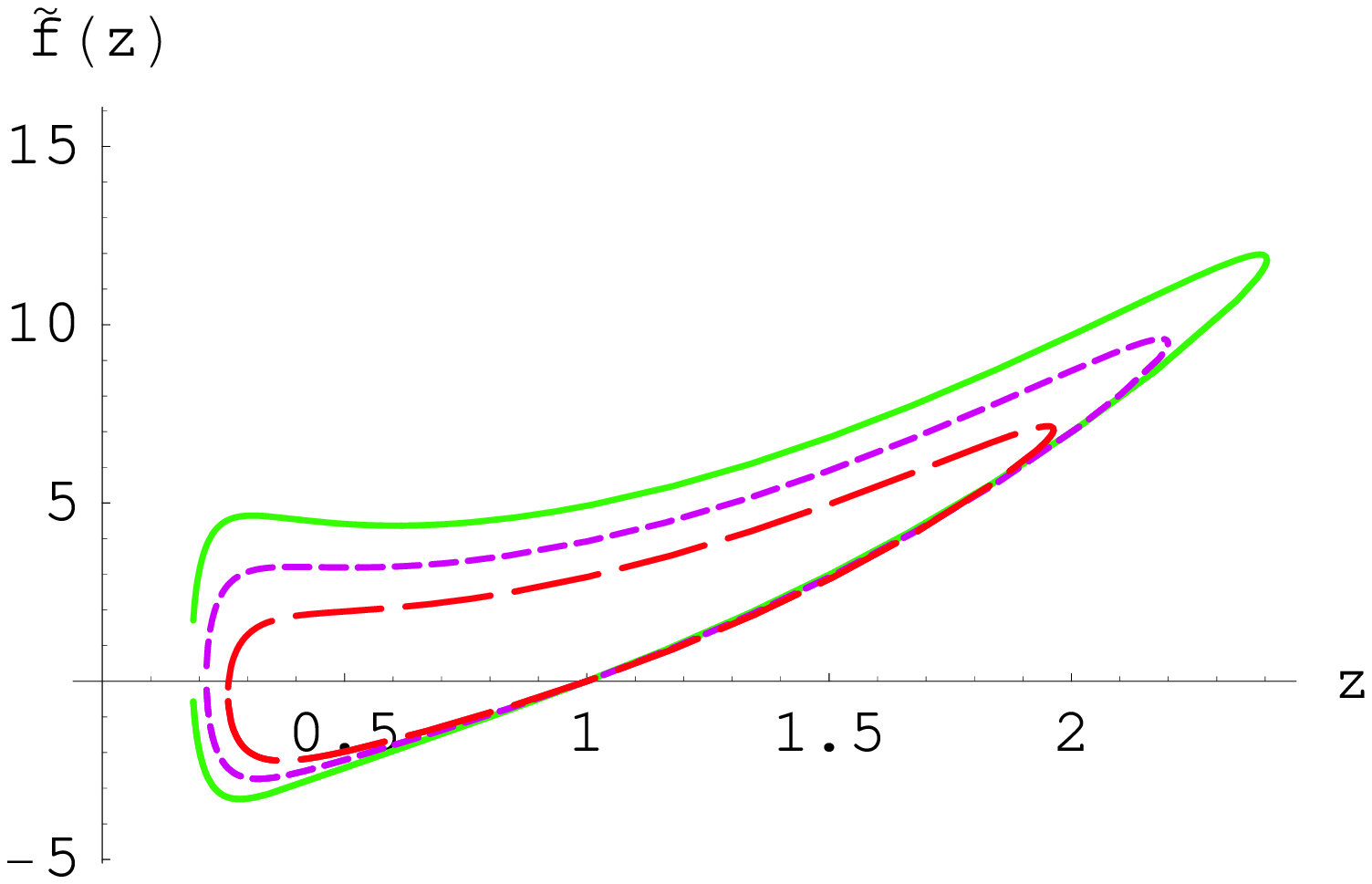,height=2.0in,width=2.9in}
\end{center}
\caption{\label{appen-fig3} As in Figure (\ref{appen-fig2}) but
with $\lambda\Z{\rm GB}=0.255$. For $\lambda\Z{\rm GB}> 1/4$, and
with an electrically charged black hole, the two branches may join
and form a closed loop.}
\end{figure}

In general, one evaluates the mass parameter $\mu$ at the black
hole event horizon ($r=r\Z{+}$) where $f(r)=0$. However, one
should note that for a given $\mu$, $f(r)$ vanishes at $r=r_+$
only for the $\kappa=-1$ branch (see figure \ref{appen-fig1}). In
that sense, the $\kappa=+1$ branch does not represent a black hole
solution. More importantly, the ADM formula, as given in
(\ref{mass-formula}), is valid only for the Einstein branch
($\kappa=-1$). For this branch, the metric potential $f(r)$ can be
written as
\begin{equation}
\tilde{f}(z)\equiv f(r)\frac{L^2}{r_+^2} = \tilde{\epsilon}
+\frac{z^2}{2\lambda\Z{\rm GB}} \left[1+\kappa
\sqrt{1-4\lambda\Z{\rm GB}+\frac{4\lambda\Z{\rm
GB}}{z^{d-1}}\left(1+\tilde{\epsilon}+\lambda\Z{\rm GB}
\tilde{\epsilon}^2+\tilde{Q}^2\left(1-z^{3-d}\right)\right)}\right]
\end{equation}
where $z\equiv r/r_+$, $x\equiv r_+/L$, $\tilde{\epsilon}\equiv
\epsilon/x^2$ and $\tilde{Q}^2\equiv  Q^2 L^2 r_+^{4-2d}$
(electric charge) or $\tilde{Q}^2\equiv  - P^2 L^2 r_+^{4-2d}$
(magnetic charge). Although the value $\lambda\Z{\rm GB}> 1/4$ may
be allowed, especially, for the positive root solution, such a
large coupling is generally not allowed for magnetically charged
black holes, so we rule out this possibility. We also note that,
with a suitable choice of $\lambda\Z{\rm GB}$, and especially, for
an electrically charged AdS-GB black hole, the $\kappa=\pm 1$
branches of solutions may join together without developing a
metric singularity (see figures \ref{appen-fig2} and
\ref{appen-fig3}).

In a more general setup, for instance, in Einstein-Gauss-Bonnet
(EGB) braneworld models~\cite{Ish-Cho-01c}, the Einstein branch
($\kappa=-1$) refers to a pure EGB gravity in higher dimensions
($d\ge 5$), while the GB branch ($\kappa=+1$) refers to matter
being confined to a 3-brane and gravity leaking into extra
dimensions. For the $\kappa=+1$ branch, a gravitational source
(like the mass term $\mu$) could reside on the brane. That is why
this branch has no Einstein limit. In that sense, only matter-free
case could be relevant for the $\kappa=+1$ branch, for which the
background is always AdS. With a nonzero $\mu$, the $\kappa=+1$
branch can produce a repulsive gravity at small distances (see
also~\cite{Padilla08b}). As far as the black hole solution is
concerned, the Einstein branch ($\kappa=-1$) is perhaps the only
relevant one. On the other hand if we have a brane-bulk system,
then the GB branch ($\kappa=+1$) also becomes relevant,
especially, for a matter-free bulk.

We end this appendix by exploring the possibility that GB gravity
connects the classical Einstein gravity to something pertinent (or
expected) in a consistent quantum gravity theory. To this end, let
us first note that in $d$-dimensions,
\begin{equation}
R\Z{\rm GB}^2 = x^{d-2} \frac{d}{dz}\left(z^{d-5}
\varphi^2\right)^\prime + {\rm surface ~terms},
\end{equation}
where $\varphi\equiv (\tilde{\epsilon}- \tilde{f}(z))$. In $d\ge
5$ dimensions, the GB contribution becomes dominant as singularity
is approached ($z\to 0$), which may therefore help in regularizing
black hole solution by weakening the singularity at $z=0$ (see
also~\cite{Dadhich:2005}). Comparing the behavior of solutions in
$d=5$ and $d=6$, we will find that the $d=5$ case gives several
desirable features including that the metric function $f(r)$ and
its derivative both remain finite and regular at $r=0$. The GB
term has the important feature of weakening of singularity which
is in consonance with what is expected in quantum gravity.

While the GB term appears as the leading correction to the
effective low-energy action of the heterotic string theory,
quantum gravity effects could easily induce infinite number of
corrections, including non-local terms. The latter endow some
unphysical (ghost) modes. To avoid any ghost degrees of freedom,
one could allow the quadratic curvature corrections in a
Gauss-Bonnet combination. In the holographic framework, any ghost
field, if exists, is expected to decouple in the large 't Hooft
coupling limit, $g_{YM}^2 N_c\to \infty\gg 1$. This is reminiscent
of the fact that in the full string theory there is (or should be)
no such problem.

In terming the GB gravity as intermediatory between classical and
quantum gravitities, our main point is that what is expected of QG
is to remove or smoothen the singularity at $r=0$. One would
expect a consistent quantum gravity theory also to remove
divergence at $r=0$ even of curvature too. The GB contribution has
a right effect in removal of singularity - at least it weakens the
singularity. This is accomplished through a change in
$r$-dependence of the metric and its derivative as $r\to 0$. For
brevity, let us consider the $d=5$ case. Note that as $r\to 0$,
\begin{equation}
f(r) \to \epsilon \mp
\frac{\sqrt{\mu\alpha}}{\alpha}+\frac{r^2}{2\alpha} \mp {\cal
O}(r^4),
\end{equation}
where $\alpha\equiv \lambda_{GB} L^2$. This shows that for either
branch there is no $\lambda\Z{\rm GB}\to 0$ limit for small $r$.
This is true also in the $Q^2\ne 0$ case, for which
\begin{equation}
 f(r)\to
\epsilon \mp \frac{\sqrt{-\alpha Q^2}}{\alpha r} \pm
\frac{\mu\sqrt{-\alpha Q^2}}{2\alpha Q^2}\, r+
\frac{r^2}{2\alpha}\mp {\cal O}(r^3).
\end{equation}
The existence of a black hole horizon generally requires $\mu
> \lambda\Z{\rm GB} L^2$, indicating that $\lambda\Z{\rm GB}$ is always
non-ignorable in a small $r$ limit. Especially, in the $Q^2=0$
case,
\begin{equation}
f^\prime \sim \frac{r}{\lambda\Z{\rm GB} L^2}.
\end{equation}
With $|\lambda\Z{\rm GB}|>0$, there is a change in radial
dependence from $1/r^3$ to proportional to $r$, which is analogous
to a situation in loop quantum gravity. A similar behavior happens
in quantum cosmology. We believe that a final QG theory will have
a unique GB or rather a more general Lovelock realization because
only then it would have a valid classical limit for which a proper
initial value problem could be defined. The requirement of a
unique evolution from an initial data might also single out some
significance of extended Lovelock gravity, where the GB term is a
natural next step of iteration of self interaction.

\end{document}